\documentclass[%
reprint,
amsmath,amssymb,
aps,
prb,
]{revtex4-1}
\usepackage{graphicx}
\usepackage{dcolumn}
\usepackage{bm}
\usepackage{color}
\usepackage{subfigure}
\usepackage{hyperref}
\usepackage[mathlines]{lineno}

\begin{document} 

\title{Multi-scale simulations of electron and ion dynamics in self-irradiated silicon} 

\author{Cheng-Wei Lee}
\affiliation{Department of Materials Science and Engineering, University of Illinois at Urbana-Champaign, Urbana, IL 61801, USA}

\author{James A. Stewart}
\affiliation{Center for Integrated Nanotechnologies, Sandia National Laboratories, Albuquerque, NM 87185, USA}

\author{Stephen M. Foiles}
\affiliation{Computational Materials and Data Science Department, Sandia National Laboratories, Albuquerque, NM 87185, USA}

\author{R\'emi Dingreville}
\affiliation{Center for Integrated Nanotechnologies, Sandia National Laboratories, Albuquerque, NM 87185, USA}

\author{Andr\'e Schleife}
\email{schleife@illinois.edu}
\affiliation{Department of Materials Science and Engineering, University of Illinois at Urbana-Champaign, Urbana, IL 61801, USA}
\affiliation{Materials Research Laboratory, University of Illinois at Urbana-Champaign, Urbana, IL 61801, USA}
\affiliation{National Center for Supercomputing Applications, University of Illinois at Urbana-Champaign, Urbana, IL 61801, USA}


\begin{abstract}
The interaction of energetic ions with the electronic and ionic system of target materials is an interesting but challenging multi-scale problem and understanding of the early stages after impact of heavy, initially charged ions is particularly poor.
At the same time, energy deposition during these early stages determines later formation of damage cascades.
We address the multi-scale character by combining real-time time-dependent density functional theory for electron dynamics with molecular dynamics simulations of damage cascades.
Our first-principles simulations prove that core electrons affect electronic stopping and have an unexpected influence on the charge state of the projectile.
We show that this effect is absent for light projectiles, but dominates the stopping physics for heavy projectiles.
By parameterizing an inelastic energy loss friction term in the molecular dynamics simulations using our first-principles results, we also show a qualitative influence of electronic stopping physics on radiation-damage cascades.
\end{abstract}

\maketitle 

\section{\label{sxn:intro}Introduction}

Interesting fundamental science and practical applications associated with the interaction of particle radiation with materials attract the attention of researchers for more than 100 years \cite{Rutherford:1911}.
More recently, projectile ions with high kinetic energies are of special interest, since for these the dominating energy-loss physics changes from inelastic electron-ion interaction (i.e., electronic stopping) to elastic ion-ion scattering (i.e., nuclear stopping) as the projectile decelerates to a full stop in the target \cite{Ziegler_1977, Averback_1997, Ziegler_2010}.
This transition affects the formation of point defects and displacement damage structures within the target material, with broad implications for the properties of the irradiated materials.
In particular, defects created by fast incident projectiles  modify a material's electrical and mechanical properties and, thus, operational performance\cite{Plummer_2000, Streetman_2006}.
This has profound consequences for practical applications with highest societal importance such as nuclear energy and safety \cite{Milbrath_2008,Zinkle_2018}, medical physics therapy \cite{Schardt_2010, Durante_2010}, space-based microelectronics \cite{Yamaguchi_2001}, and fundamental-research laboratories \cite{Ward_2006, Krasheninnikov_2010,Kim_2012,Lindquist_2012}.
Understanding the many underlying exciting fundamental questions creates an immense need for predictive modeling.

\begin{figure}
\centering
\includegraphics[width=0.9\columnwidth]{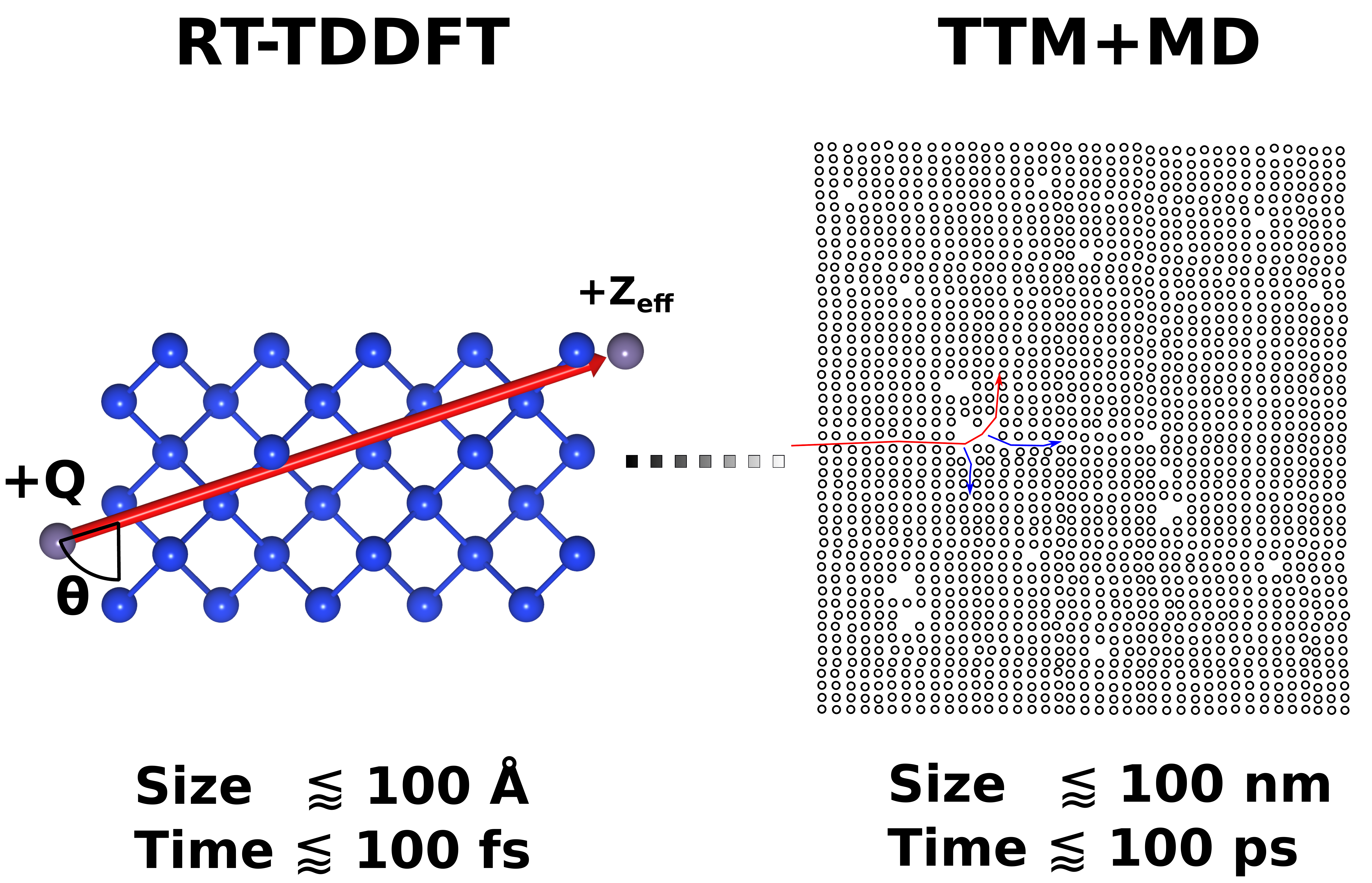}
\caption{\label{fig:multi}
Multi-scale nature of projectile-target interactions.
Typical length ($\sim$ 1 -- 100 nm) and time scales ($\lesssim$ 100 ps) associated with irradiation processes~\cite{Averback_1997,Dudarev_2013, Darkins_2018},
responsible for an inherent multi-scale nature in coupling stopping power and damage production.
This renders an accurate and quantitative description of the combined phenomena a challenge.
Here we combine real-time time-dependent density functional theory (RT-TDDFT) and the two-temperature model (TTM) in molecular dynamics (MD) simulations.
}
\end{figure}

At the same time, understanding effects of energetic ions on materials from a theoretical or computational description is a challenging multi-scale problem (see Fig.\ \ref{fig:multi}).
Generally, the energy transfer from projectile ions into targets is described by the stopping power $S$, a compound quantity that is defined as the energy loss $dE$ of the projectile per penetration depth $dx$,
\begin{equation}
  \label{eqn:estopping-power}
  S(E)= dE(x)/dx,
\end{equation}
and has units of force \cite{Correa_2018, Was_2016}.
Due to the aforementioned multi-scale character---attributed to the involvement of electronic and ionic system of the target---projectile-target interactions and, thus, $S$ depend on the target material, the projectile kinetic energy as well as charge state and impact angle of the projectile.
Modeling the extent of defects and damage in the target requires understanding these details of energy deposition processes and their complicated dependencies.
This triggered a strong interest in the computational materials science community to accurately predict $S$ and also parameters such as penetration depth of incident particles, type of defects formed, and size of the defect structures created during radiation damage events.

Typically, existing models simplify the problem by decoupling different time scales via the projectile velocity:
Upon impact of energetic (swift) ions with velocities greater than the Bohr velocity $v_{B}$=$e^{2}/\hbar$, non-adiabatic interactions excite the electronic system of the target (see left panel of Fig.\ \ref{fig:multi}), accompanied by intriguing femto- to pico-second equilibration and relaxation dynamics.
During this initial stage, the predominant energy-loss mechanism of the projectile is electronic stopping ($S_e$), whereas nuclear stopping can largely be neglected.
Common analytic models for $S_e$ include Lindhard-Winther theory \cite{Lindhard_1964}, which describes the electronic excitations via linear plasmonic response.
In this model electronic stopping depends quadratically on the effective charge $Z_\mathrm{eff}$ of the projectile with velocity $v$, and increases with the electron density $n$ of the target material,
\begin{equation}
\label{eq:lindhard}
-\frac{dE}{dx}=\frac{4\pi}{m}\left(\frac{{Z_\mathrm{eff}} e^{2}}{v}\right)^{2} n L(v,n).
\end{equation}
$L(v,n)$ is the stopping number, describing the linear plasmonic response of the uniform electron gas, which can be calculated as a double integral in energy and momentum of the energy loss function using, e.g., the random-phase approximation \cite{Schleife_2015, Correa_2018}.
Alternatively, electronic stopping can be calculated using scattering theory formulation.
Early versions, such as the Bethe formulation \cite{Bethe_1930}, treat the scattering events (semi-)classically.
Along with the development of quantum mechanics, many-body effects like electronic screening were included;
one example is the Lindhard-Scharff theory \cite{Lindhard_1961}.
At low velocities, a first-order approximation of this theory shows that $S_{e}$ is linearly proportional to the incident particle velocity, similar to Lindhard-Winther theory for low velocities.
The approximate validity of this model in the low-velocity regime originally motivated using a linear friction term in the two-temperature model \cite{Darkins_2018}, to incorporate electronic stopping into molecular dynamics (MD) simulations.
However, we note that analytical versions of both approaches \cite{Bethe_1930, Lindhard_1961, Lindhard_1964} require the effective charge of the projectile and the number of electrons involved in the electronic excitation as external parameters \cite{Correa_2018}.
In going beyond analytical models, an important milestone was enabled by density functional theory (DFT), and takes into account the self-consistent screened potential for describing stopping of slow ions in a homogeneous \cite{Echenique_1981, Echenique_1986} and inhomogeneous \cite{Nagy_1988} electron gas. 
Later work illustrated the importance of \emph{dynamic} many-body exchange-correlation effects for accurate predictions of electronic stopping of slow ions even in the zero-velocity limit \cite{Nazarov_2005,Nazarov_2007}. 

As the swift ion loses energy and slows down in the target material to $v \ll v_{B}$,
collisions with nuclei become the dominant energy-loss mechanism, leading to the formation of displacement cascade structures.
The corresponding ion dynamics in this low-energy nuclear stopping regime can be modeled on an atomistic level using classical or ab-initio (Born-Oppenheimer) MD simulations.
Alternatively, Monte-Carlo techniques and the binary collision approximation can be used, as implemented in the open-source package ``Stopping and Range of Ions in Matter'' (SRIM)  \cite{Ziegler_1985, Ziegler_2010}.
However, including electronic stopping effects in these simulations, to incorporate the multi-scale character, is a challenge.
As discussed above, this can be accomplished to zeroth order using a linear friction term, fitted to electronic stopping results predicted by more accurate methods such as Lindhard-Winther theory \cite{Lindhard_1964}.
Alternatively, Bethe-Bloch theory (with additional corrections to account for projectile charge state and relativistic effects and fitted to experiment) has proven very successful to account for electronic stopping \cite{Ziegler_1999, Ziegler_2010}, e.g.\ in SRIM.

Unfortunately, such an approach does not achieve a full multi-scale description of the interaction of swift ions with target materials, and, thus, effectively hampers detailed and precise understanding:
In particular, one of the disadvantages of the Bethe-Bloch or Lindhard-Winther approaches is the assumption of an amorphous target material, which entirely neglects
any impact of the crystalline structure and local electron density, e.g., of channeling vs.\ off-channeling projectile trajectories, on electronic stopping.
In order to achieve the necessary accuracy of MD simulations to precisely describe radiation damage, e.g.\ for ion strikes in crystalline samples \cite{Peltola_2006, Race_2012}, it is essential to incorporate a structural or directional dependence.

In principle, this issue naturally lends itself to first-principles calculations that explicitly consider electron-electron and electron-ion interactions, and crystal structure.
To this end, real-time time-dependent density functional theory (RT-TDDFT) has recently been used successfully to predict electronic stopping power for a diverse range of target materials and incident projectiles, as summarized in Ref.\ \onlinecite{Correa_2018} and references therein.
However, the majority of RT-TDDFT stopping power studies focused only on light incident projectile ions, such as protons or alpha particles and, unfortunately, only few studies consider heavy ions ($Z>2$).
The few available studies report much more intricate stopping physics for heavy ions:
Lim \emph{et al.}\cite{Lim_2016} studied a Si projectile with kinetic energies between 1 eV and 100 keV penetrating bulk Si along a $\langle 001 \rangle$ channel. 
They showed delicate band structure effects in the low kinetic energy regime.
Ojanper{\"a} \emph{et al.}\cite{Ojanpera_2014} showed that projectile core electrons are critical for accurate predictions of electronic stopping for heavy ions penetrating graphene.
Ullah \emph{et al.}\cite{Ullah_2018} showed that for self-irradiated Ni, core electrons of the projectile are more important for electronic stopping than those of target atoms and they proved the importance of core electrons of target atoms especially in the high kinetic energy regime.
However, even these studies of heavy projectiles are limited to channeling trajectories and restricted to weakly charged projectile ions.
In particular, they do not attempt to explain charge equilibration of projectiles, or the connection of empty projectile core states and electronic stopping, both of which are expected to become important for strongly charged heavy projectiles.

In addition, as discussed above, to achieve accurate multi-scale simulations of radiation damage (see Fig.\ \ref{fig:multi}), it is necessary to propagate fundamental understanding of electronic stopping experienced by energetic projectiles beyond electronic length and time scales \cite{Tamm:2019,Caro:2019}.
This has been demonstrated to be an essential component for accurate ion dynamics \cite{Zarkadoula:2018, Lee_2019} and radiation-damage cascade simulations using molecular dynamics (MD).
While the influence of electronic stopping is typically ignored entirely, in some cases it was incorporated through various supplements to the MD equations of motion, and, most commonly, through an inelastic energy loss (IEL) friction term \cite{Nordlund_1995, Cai_1996, Zhong_1998, Tian_2018, Stewart_2018}.
Recently, Sand \emph{et al.}\cite{Sand_2019} combined electronic stopping results from RT-TDDFT with IEL based MD to predict the tungsten ion range in tungsten targets and reported good agreement with experiment.
An extension of this IEL approach, called the two-temperature model (TTM), considers not only electronic stopping effects via a frictional drag force but also electron-phonon coupling and electronic heat transfer to account for ``cold electrons'' moving through thermally spiked regions and becoming excited.
This electron-phonon coupling is described through the use of a stochastic force term to allow energy transfer between the atomic lattice and electronic subsystems~\cite{Caro_1989, Duffy_2007, Rutherford_2007, Darkins_2018}.

In this work we use a multi-scale modeling approach to investigate the role of electronic stopping for radiation damage in bulk silicon.
This material is of great technological interest and is widely studied, making it a well-suited testbed for this work.
We use RT-TDDFT simulations to account for crystal structure (channeling vs.\ off-channeling), core electrons, and charge state of proton and silicon (light vs.\ heavy) projectiles when predicting electronic stopping power.
We show that electronic stopping depends on the initial projectile charge for channeling projectiles.
By comparing the limiting cases of initially neutral Si$^{+0}$ and initially highly charged Si$^{+12}$ projectiles our simulations prove that core electrons of the target atoms play a crucial role for charge equilibration of initially highly charged ions.
We subsequently integrate these RT-TDDFT results for electronic stopping into MD simulations of single-ion strike radiation-damage via the IEL and TTM approaches.
For both models the influence of the proportionality constant between incident projectile velocity and electronic stopping power, $\gamma_{e}$, needs to be understood.
To this end, we compare data from the integrated multi-scale approach with electronic stopping from RT-TDDFT against MD results without electronic stopping effects and MD results with electronic stopping fitted to SRIM predictions.
Using these results we illustrate the importance of correctly capturing electronic stopping effects within MD and elucidate the strong influence on the resulting damage structure.
Our findings provide high-accuracy first-principles insight into radiation damage over multiple length and time scales and we envision an integration of simulation techniques as presented here to be instrumental for achieving the full potential of radiation-material interactions.

\section{\label{sxn:method}Computational Methods}

To study multi-scale processes occurring during cascade evolution in bulk silicon, we carried out calculations combining RT-TDDFT and classical MD.
The RT-TDDFT calculations have a dual purpose:
On the one hand, they provide necessary calibration of the proportionality constant $\gamma_e$ used in the MD simulations.
On the other hand, RT-TDDFT simulations provide physical insight into the role of atomic structure, charge state, and core electrons on electronic stopping power.
The MD simulations of single ion-strike radiation damage incorporate electronic stopping power via the inelastic energy loss (IEL) approach using a quadratic formulation of the stopping power, which provides a better representation of the electronic stopping force as compared to the linear description of Ref.\ \onlinecite{Lindhard_1961}.

\subsection{\label{sec:gs}Electronic ground state}

First, we use the open-source Qb@ll code~\cite{Schleife_2014,Draeger_2017} to perform ground-state density functional theory (DFT)~\cite{Hohenberg_1964, Kohn_1965} calculations for bulk silicon in the diamond structure (space group: Fd$\bar{3}$m).
We use the local-density approximation (LDA) \cite{Ceperley_1980} to describe exchange and correlation and the electron-ion interaction is described by Hamann-type norm-conserving pseudo-potentials \cite{Hamann_1979}.
To explicitly study the influence of semi-core states, we compare a pseudo-potential with four valence electrons (Si\,$3s$, Si\,$3p$), generated using the modification by Vanderbilt \cite{Vanderbilt_1985}, to one with 12 valence electrons (Si\,$3s$, Si\,$3p$, Si\,$2s$, and Si\,$2p$), using the modification by Rappe \emph{et al.}\ \cite{Rappe_1990}.
We note that for computational reasons we restrict all simulations for channeling \emph{proton} projectiles to the pseudo-potential with four valence electrons, but use the one with 12 valence electrons for off-channeling projectiles.
This is motivated by earlier work \cite{Schleife_2015}, which showed that semi-core electrons do not contribute to electronic stopping for light, channeling projectiles.
Kohn-Sham states are expanded into a plane-wave basis with cutoff energies of 1360 eV and 2450 eV for four and 12 valence electrons, respectively.
These numerical parameters allow us to compute total energies converged to within 5 meV/atom.
The Brillouin zone of the 216-atom supercell studied here is sampled using only the $\Gamma$ point.
Furthermore, we obtain relaxed atomic coordinates from fits to the Murnaghan equation of state \cite{Murnaghan_1944} and minimization of Hellman-Feynman forces to below 5 meV/\AA.
We find a lattice constant of 5.37 \AA, which slightly underestimates the experimental value of 5.43~\AA ~\cite{Dargys_1994}.

\subsection{\label{sec:TDDFT}Real-time dynamics and electronic stopping}

We use real-time time-dependent DFT (RT-TDDFT) and the Ehrenfest molecular dynamics approach to describe electron-ion dynamics~\cite{Ehrenfest_1927, Marx_2009, Runge_1984} within the Qb@ll code \cite{Schleife_2014, Draeger_2017}.
In our electron dynamics simulations, we use adiabatic LDA to describe exchange and correlation. 
Although LDA has a well-known problem of underestimating the electronic band gap, it has been shown that electronic band gaps have little effect on electronic stopping for large enough projectile kinetic energies, e.g. KE $>$ 3 keV for Si projectiles in Ref. \onlinecite{Lim_2016}, which is also the regime studied in the present manuscript.
Ground-state Kohn-Sham wave functions from DFT serve as initial condition for the real-time propagation of electronic states.
Upon ion irradiation the time-dependence in the Hamiltonian is driven by fast proton and silicon projectiles moving through bulk silicon.
Time-dependent Kohn-Sham equations are propagated in real time using a fourth-order Runge-Kutta integrator~\cite{Schleife_2012}.
Integration time steps were chosen such that the influence on electronic stopping is less than 0.1\,\% when the time step is reduced by a factor of two.
This leads to a time step of 14.5$\times$10$^{-3}$ atomic units (at.\ u.), corresponding to 0.35 as, in most of the presented cases;
only for silicon projectiles with kinetic energies larger than 6.27 MeV a smaller time step of 7.25$\times$10$^{-3}$ at.\ u.\ (i.e., 0.175 as) is used.
Non-adiabatic electron-ion coupling is taken into account via Hellman-Feynman forces, computed from the time-dependent electron density~\cite{Ehrenfest_1927,Marx_2009}.
In our simulations we describe an initially neutral Si projectile by including the projectile Si atom in the initial DFT ground-state calculation.
The ionized projectile is simulated using a DFT ground state of ideal bulk silicon to which we add the projectile ion and immediately start real-time propagation without any further electronic optimization \cite{Schleife_2015}.

We compute instantaneous electronic stopping $S(x)$ using Eq.\ \eqref{eqn:estopping-power} and the energy increase of the electronic system as the projectile penetrates the material.
We use Ehrenfest MD to study projectiles traversing channeling directions in bulk silicon.
Along the $\langle 001 \rangle$ direction, the 216-atom supercell used in this work has three lattice periods.
The lengths of the lattice periods along the $\langle 011 \rangle$ and $\langle 111 \rangle$ directions are increased by factors of $\sqrt{2}$ and $\sqrt{3}$, respectively.
For protons moving through $\langle 001 \rangle$, $\langle 011 \rangle$, and $\langle 111 \rangle$ channels, we average the instantaneous stopping $S(x)$, computed using Ehrenfest molecular dynamics, by integrating over two lattice periods.
This is to reduce onset effects by discarding the first half lattice period of the simulation (see Supplemental Material at [URL will be inserted by publisher] for the detailed discussion).
Additionally, by discarding the last half lattice period of the simulation, we remove effects of excited electrons that re-enter the simulation cell due to periodic boundary conditions \cite{Schleife_2015, Correa_2018}.
For Si projectiles, the average stopping is instead calculated using the slope of a linear regression fit to the energy gain vs.\ displacement curve for the same two lattice periods.
Compared to the approach above, this reduces the fitting error when the oscillation magnitude is large.
Alternatively, Quashie \emph{et al.}\ \cite{Quashie_2016} showed that using an oscillatory fit, $y=a+bx+A\cos(kx+\phi)$, further reduces the fitting error for projectiles with low velocity.

Off-channeling conditions are studied using RT-TDDFT simulations for a projectile on a random trajectory through the silicon crystal, as described in Ref.\ \onlinecite{Schleife_2015}.
We use a pseudo-random number generator to generate a random direction, along which the projectile moves through the lattice. 
We verified that the trajectory is dissimilar from any lattice channel and that the simulations are long enough to obtain converged results. 
All atoms of the target material are fixed on their equilibrium lattice sites to avoid numerical issues caused by very short separations between the projectile and target atoms (see details in Ref.\ \onlinecite{Schleife_2015}). 
Average electronic stopping in this case is calculated using the slope of a linear regression fit to the energy gain vs.\ displacement curve \cite{Schleife_2015}.
Additionally, due to the projectile charge dynamics  observed for Si ions in this work, we remove the data before the projectile charge reaches equilibrium, to ensure that the calculated electronic stopping is indeed for the equilibrium charge state.
Specifically, the data before projectile trajectory length of 34 and 5.3 \AA \ was removed from the fitting for initially ionized and neutral Si projectiles, respectively.
Initially, this result is sensitive to the trajectory length; however, we were able to find convergence for trajectories of approximately 400 and 200 \AA \ for proton and Si projectiles, respectively. 
Computational cost prevents simulation of such long trajectories for the Si projectile with low kinetic energy ($\leq$ 2.79 MeV), resulting in an overestimation of electronic stopping by about 5\,--\,15 \% (see Supplemental Material at [URL will be
 inserted by publisher] for the details).

\subsection{\label{ssxn:ddexc6}Projectile charge state}

We use the density-derived electrostatic and chemical (DDEC6) method \cite{Manz_2016}, a modern charge-decomposition scheme, to compute the charge associated with the projectile as it travels through the target.
By taking atomic orbitals into account when assigning electron density to all atoms, given a total charge density, the DDEC6 approach goes beyond a mere spatial decomposition, as done in Bader or Voronoi analysis.
While DDEC6 was derived for electronic ground-state densities \cite{Manz_2016}, we expect a small error when applying it in this work, since only a small fraction of the total number of electrons is in excited states.

\subsection{\label{ssxn:md-methods}MD simulation of displacement damage}

Following the simulation methodology of Ref.\ \onlinecite{Stewart_2018}, we utilize the open-source Large-scale Atomic/Molecular Massively Parallel Simulator (LAMMPS) atomistic code~\cite{Plimpton_1995} to perform simulations of single isolated PKA displacement cascades with a recoil energy of 20 keV and a recoil trajectory towards the center of a bulk crystalline Si sample.
The kinetic energy of 20 keV used here corresponds to the experimentally relevant regime when 46 keV Au projectiles impact Si target materials and, at the same time, provides a reasonable MD domain size, allowing us to perform the many simulations for this work.
The displacement cascade development in these simulations is accommodated by using a cubic simulation domain, containing 6,229,504 atoms, that is $50~\textrm{nm} \times 50~\textrm{nm} \times 50~\textrm{nm}$ with periodic boundary conditions in each direction.
We use the 3-body Tersoff interatomic potential to describe all interatomic interactions~\cite{Tersoff_1988}.
In addition, nuclear stopping is taken into account via the Ziegler-Biersack-Littmark (ZBL) universal screening function to correct the Tersoff interatomic description at very short interatomic separations, which are readily created during collision cascades~\cite{Ziegler_1985}.
We perform the cascade simulations at a temperature of 300 K.
As such, prior to PKA initiation, the crystalline Si structure is equilibrated to 300 K for 3.3 ps using a constant-temperature, constant-pressure ensemble (NPT).
The displacement cascade is initiated by randomly choosing a PKA between 11~\AA\ and 55~\AA\ from the domain boundary and assigning a velocity corresponding to a kinetic energy of 20 keV and a random direction into the bulk Si structure (see Fig.\ \ref{fig:defect-shape}).

In our simulations, the displacement cascade formation and evolution are performed at 300 K using a micro-canonical ensemble (NVE) with an adaptive time step ($10^{-7}~\rm{ps} \leq \Delta t \leq10^{-3}~\rm{ps}$) for 85,000 time steps, allowing for a maximum 85 ps cascade simulation.
Note that the observed simulation times for the majority of MD simulations presented are approximately 10\,--\,15 ps less than this maximum due to the need for smaller time steps in the very early stages of cascade formation.
We apply a 300 K Langevin thermostat (i.e., stochastically damped equations of motion) within 11~\AA\ of all boundaries of the simulation domain to allow for any excess kinetic energy associated with the shock wave introduced by the recoiling PKA to be dissipated, thus mimicking dissipation that would occur within an infinite medium.
The displacement cascade simulation is then followed by an annealing period at 300 K for 15 ps by switching to an NPT thermostat to allow for the defects to recover and form a thermodynamically stable defect structure.
We identify and count the point defects (i.e., vacancies and interstitials) generated during cascade formation by using a Wigner-Seitz cell analysis of the atomic positions in the damaged Si structure with respect to the undamaged Si structure~\cite{Stukowski_2010}.
With this method, a lattice site with an empty Wigner-Seitz cell is marked as a vacancy while a Wigner-Seitz cell with two or more atoms is marked as an interstitial.

We performed and compare three different types of displacement cascade simulations:
(i) neglecting electronic stopping power entirely, (ii) approximating electronic stopping power via a frictional drag force as a function of atomic velocity that only describes inelastic energy loss (IEL), and (iii) approximating electronic stopping via the same frictional drag force as (ii), but using the two-temperature model framework that allows for electron-ion interactions and energy transfer between the atomic and excited electronic subsystems \cite{Caro_1989, Duffy_2007, Rutherford_2007}.
In order to obtain statistically meaningful results, 10 simulations are performed for each case of electronic stopping method and stopping power fitting data (i\@.e\@., SRIM and RT-TDDFT).

In the case of the IEL method, when incorporating electronic stopping effects using a friction term to only describe energy loss, the governing equations of motion are given as~\cite{Caro_1989}
\begin{equation}
\label{eqn:iel-drag}
m_{i}\frac{\partial \textbf{v}_{i}}{\partial t} = \textbf{F}_{i}(t) - \gamma_{e}\textbf{v}_{i}~,
\end{equation}
where $m_{i}$ and $\textbf{v}_{i}$ are the mass and velocity of atom $i$, $\textbf{F}_{i}(t)$ is the force acting on atom $i$ at time $t$ due to its interactions with atoms in its local environment, and $\gamma_{e}$ is the friction coefficient due to electronic stopping above a cutoff velocity, $v_{\textrm{cut}}$.
In the TTM method~\cite{Duffy_2007, Rutherford_2007}, to account for electron-ion interactions and the transfer of energy lost to the electronic subsystem back into the atomic lattice subsystem, the governing equations of motions are given as
\begin{equation}
\label{eqn:ttm}
m_{i}\frac{\partial \textbf{v}_{i}}{\partial t} = \textbf{F}_{i}(t) - \gamma_{i}\textbf{v}_{i} + \widetilde{\textbf{F}}(t)~,
\end{equation}
where $\gamma_{i}$ is now a frictional force describing both the electronic stopping effects, $\gamma_{e}$, as before and electron-ion interactions, $\gamma_{p}$.
In general, $\gamma_{i} = \gamma_{p} + \gamma_{e}$ for $v_{i} > v_{\textrm{cut}}$ and $\gamma_{i} = \gamma_{p}$ for $v_{i} \leq v_{\textrm{cut}}$. $\widetilde{\textbf{F}}(t)$ is a stochastic force term with a random direction and magnitude that is a function of the electron-ion friction force coefficient, $\gamma_{p}$, and the electronic subsystem temperature, $T_{e}$.
This term allows for energy lost due to electronic stopping to be transferred back to the atomic lattice subsystem.
The electronic subsystem temperature, $T_{e}$, is described by a heat diffusion equation requiring an electron density, $\rho_{e}$, an electronic specific heat, $C_{e}$, and thermal conductivity, $\kappa_{e}$.
Furthermore, this diffusion equation contains a sink term representing the energy exchange between the electronic subsystem and the atomic lattice subsystem, and a source term representing the energy gained by the electronic subsystem (i.e., energy lost by the atomic subsystem) due to electronic stopping.

A common practice uses the cohesive energy as the choice of cutoff in the electronic stopping power for MD simulations. 
In this spirit, to prevent artificially quenching atoms in thermal equilibrium within the IEL electronic stopping description, we have truncated the electronic stopping effects between velocities corresponding to one and two times the cohesive energy of Si, where $E_{\textrm{coh}} = 4.63$~eV (or equivalently, $v_{\textrm{coh}} = 56.402$~\AA/ps). 
Based on previous RT-TDDFT calculations (c.f., Fig. 1 in Lim et al.~\cite{Lim_2016}) as well as our own data in Fig.\ \ref{fig:Se_Si}, this is a reasonable truncation regime because it takes into account both the relative magnitude of the stopping power with respect to the cohesive energy and the rate of change of the electronic stopping power as a function of the kinetic energy. 
As such, for velocities lower than the cohesive energy, no stopping force is implemented and for velocities higher than twice the cohesive energy, Eq.\ \eqref{eqn:iel-drag} is used.
In between these two velocities, the truncation is performed utilizing a quadratic polynomial given as $(v_{i}^{2} - 2v_{\textrm{coh}}^{2})/(v_{\textrm{coh}}^{2} - 2v_{\textrm{coh}}^{2})$.
In contrast, in the case of the TTM method, only a single cutoff velocity is needed to transition between purely electron-phonon and electronic stopping regimes.
From the RT-TDDFT calculations in Ref.\ \onlinecite{Lim_2016}, a reasonable value for this singular transition is found to be 6.945 eV (or $v_{\textrm{cut}}$ = 69.078~\AA/ps), which corresponds to 1.5 times $E_{\textrm{coh}}$.
We treated the electron-phonon and electronic stopping regimes as two limiting cases of the single energy transfer process.
To this end, we utilize the condition that $\gamma_{i} = \gamma_{p}$ for $v_{i} \leq v_{\textrm{cut}}$ and $\gamma_{i} = \gamma_{e}$ for $v_{i} > v_{\textrm{cut}}$.
Finally, the remaining TTM electronic stopping parameters are taken from Ref.\ \onlinecite{Jay_2017} and summarized in table~S2.

\section{\label{sxn:results}Results}

\subsection{\label{subsec:proton}Light projectiles: Stopping of protons in silicon}

\begin{figure}[!htb]
\includegraphics[width=0.9\columnwidth]{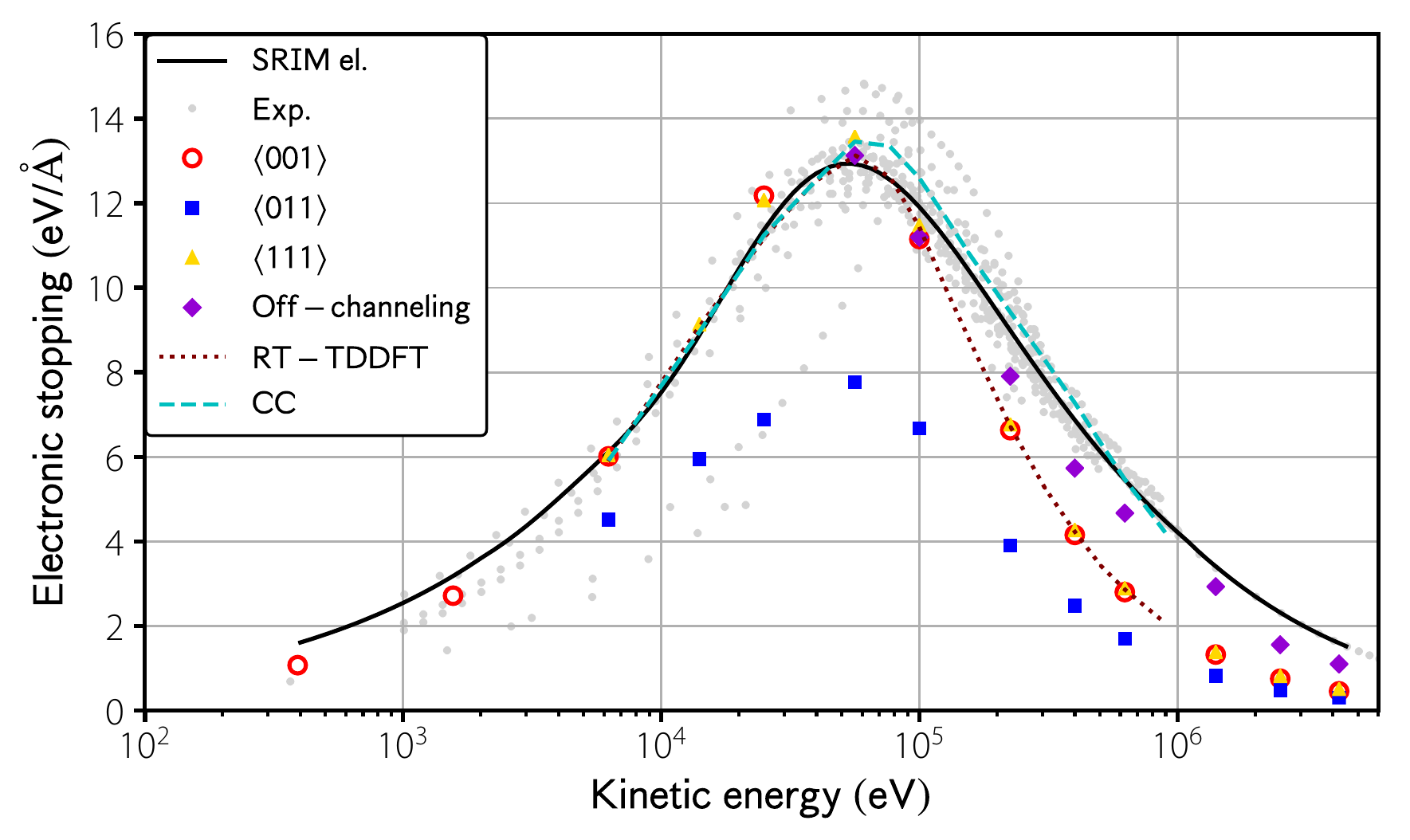}
\caption{\label{fig:Se_H}
Electronic stopping of proton projectiles.
Red circles, blue squares, and gold triangles indicate our RT-TDDFT results for $\left<001\right>$, $\left<011\right>$, and $\left<111\right>$ channels, respectively, and violet diamonds indicate off-channeling.
Gray dots are experimental values for off-channeling \cite{IAEA_data} and black solid line results from SRIM \cite{Ziegler_2010}.
Cyan dashed and maroon dotted lines are RT-TDDFT results from Ref.\ \onlinecite{Yost_2017} for $\left<001\right>$ channeling, without and with core-electron corrections (CC), respectively.
}
\end{figure}

We first establish a theoretical understanding of the femto-second electron-ion dynamics during the early stages immediately after projectile impact for stopping of light proton projectiles in Si.
In Fig.\ \ref{fig:Se_H} we illustrate results from RT-TDDFT simulations of this process for three different lattice channels, as well as one off-channeling trajectory.
Our results for low-energy ($\lessapprox$0.1 MeV) protons traveling along the $\left<001\right>$ and $\left<111\right>$ channels agree particularly well with experiment \cite{IAEA_data} and with predictions from SRIM.
Comparison with the theoretical data of Ref.\ \onlinecite{Yost_2017}, before correcting for core electrons, also shows excellent agreement for $\left<001\right>$ channeling.
While agreement for projectile kinetic energies near or lower than the electronic-stopping maximum is excellent, Fig.\ \ref{fig:Se_H} also illustrates that RT-TDDFT results for electronic stopping deviate from experiment for all channeling trajectories when the projectile kinetic energies exceeds $\approx$0.1 MeV.
Since most experiments and SRIM simulations are performed for amorphous materials or use off-channeling trajectories through crystalline targets, this deviation is attributed to off-channeling effects.

Correspondingly, Fig.\ \ref{fig:Se_H} shows that off-channeling projectiles experience the highest electronic stopping for each velocity.
Agreement between our data and SRIM is within the scatter of experimental data points up to high kinetic energies.
We also note that including core corrections, to fully account for Si\,$1s$ electrons (see data of Ref.\ \onlinecite{Yost_2017} in Fig.\ \ref{fig:Se_H}), only slightly improves agreement with experiment;
2$s$ and 2$p$ electrons, however, are crucial.
For the $\left<011\right>$ channel, experimental results of Refs.\ \onlinecite{Hobler_2006,Carnera_1978}(not shown in Fig.\ \ref{fig:Se_H}) are reported to be about 10\,\%\,--\,15\,\% lower in stopping than for off-channeling \cite{Carnera_1978}, while our data shows approximately 35\,\% lower stopping, e.g., near 20 keV.
To explain this, we note that our RT-TDDFT data constitutes a lower bound for channeling electronic stopping, since in our simulations the projectile travels exactly at the center of the channel, where the charge density is the lowest. 
This effect is more significant for the $\langle 011 \rangle$ channel, because the average charge density is much lower near the center of the channel and increases more quickly further away from it, compared to the other two channels (see Fig.\ S1B in Supplemental Material at [URL will be
 inserted by publisher] for explicit data).
Finally, while the majority of measured stopping values agree well with our RT-TDDFT data for $\left<001\right>$ and $\left<111\right>$ channels and SRIM, there is a data set \cite{Grahmann_1976} that shows much lower stopping between 10 keV and 50 keV.
This data is included in the experimental data points in Fig.\ \ref{fig:Se_H} and coincides with our results for $\left<011\right>$.
We note, however, that these experiments \cite{Grahmann_1976} were performed for \emph{off-channeling} protons and the difference from other experimental results is generally attributed to the use of a different measuring technique \cite{Carnera_1978}.

Our data also shows a pronounced dependence of electronic stopping on the specific lattice channel:
While the electronic stopping power is similar for the $\left<001\right>$ and $\left<111\right>$ channels, it is significantly smaller for a $\left<011\right>$ channel (see Fig.\ \ref{fig:Se_H}).
To explain this, we analyzed the distance between the projectile and the nearest-neighbor atoms it encounters along its path through the target (see  Fig.\ S1A in Supplemental Material at [URL will be
 inserted by publisher] for explicit data).
For the $\left<001\right>$ and $\left<111\right>$ channels this distance is only about 66\,\% of the value observed for the $\left<011\right>$ channel.
As a result, the average electron density seen by the projectile along these channels is at least a factor of three smaller for $\left<011\right>$, compared to the other two (see Fig.\ S1A in Supplemental Material at [URL will be inserted by publisher] for explicit data ).
This clearly correlates with the electronic stopping power predicted from RT-TDDFT simulations for the different channels and is consistent with previous RT-TDDFT studies \cite{Ullah_2015, Lee_2018}.
We rationalize this using the linear-response Lindhard-Winther model \cite{Lindhard_1964}, Eq.\ \eqref{eq:lindhard}.
While our first-principles data does not show the direct proportionality of electronic stopping to the electron density seen by the projectile along its trajectory, as predicted for the dilute charge-density limit (see appendix in Ref.\ \onlinecite{Ziegler_1999}), the model qualitatively explains the trends we observe.

\subsection{\label{subsec:silicon}Heavy projectile ions: Stopping for self-irradiated silicon}

\begin{figure}[!htb]
\includegraphics[width=0.9\columnwidth]{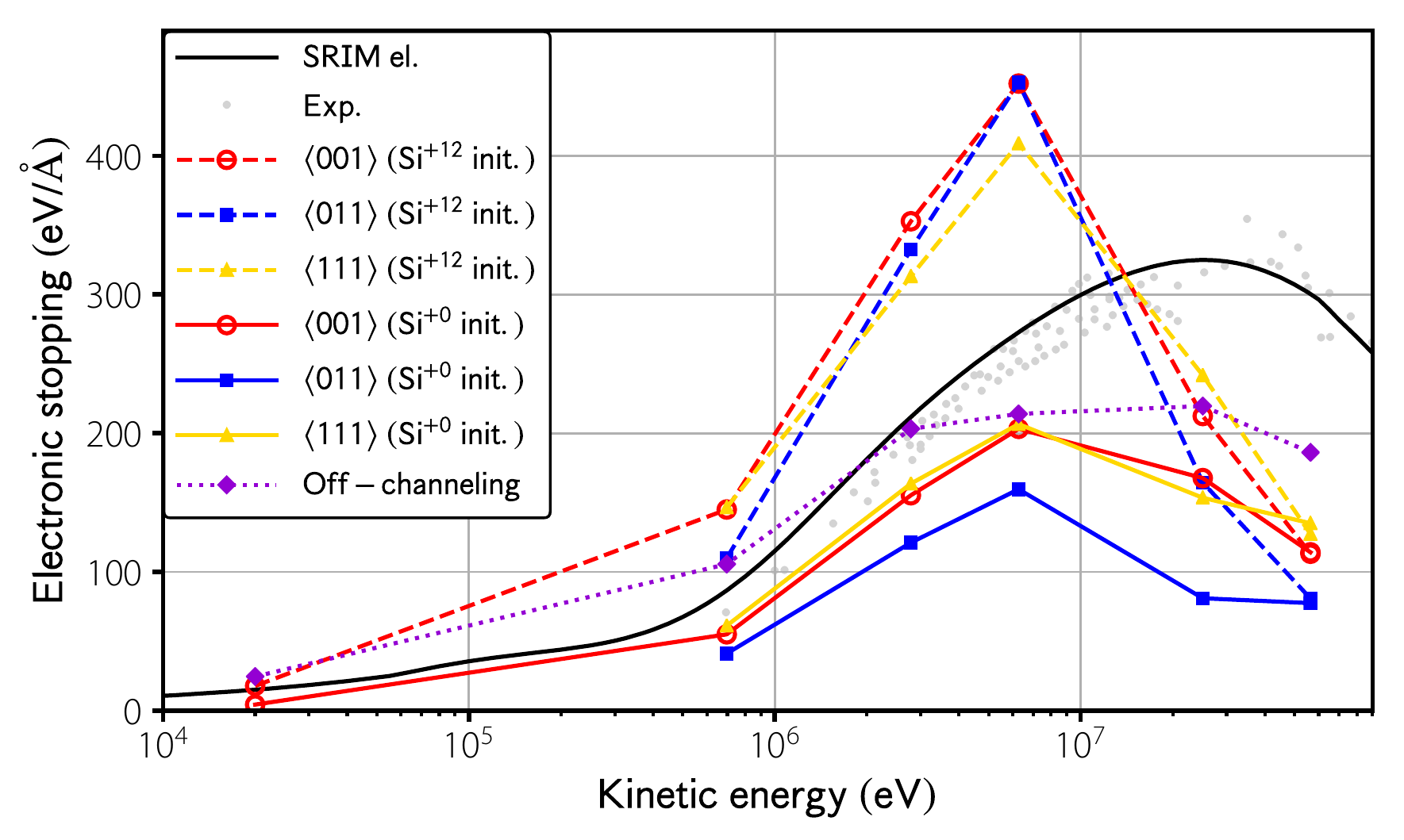}
\caption{\label{fig:Se_Si}
Electronic stopping of silicon projectiles.
Red circles, blue squares, and gold triangles indicate our RT-TDDFT results for $\left<001\right>$, $\left<011\right>$, and $\left<111\right>$ channels, respectively, and violet diamonds indicate off-channeling.
Gray dots are experimental values for off-channeling \cite{IAEA_data} and black solid line results from SRIM \cite{Ziegler_2010}.
Two different initial projectile charge states are compared using solid (neutral) and dashed lines (highly ionized).
}
\end{figure}

The picture is significantly more complicated for heavy ion projectiles:
These are important when modeling semiconductors under radiation conditions, since primary knock-on events inevitably occur in materials impacted by energetic ions.
To this end, Fig.\ \ref{fig:Se_Si} compares electronic stopping power from RT-TDDFT for self-irradiated silicon to experiment and SRIM, using the same three lattice channels we studied for protons, as well as off-channeling.
For all channeling trajectories, Fig.\ \ref{fig:Se_Si} shows that electronic stopping depends on the initial charge state of the silicon projectile.
This striking behavior is contrary to what we observed for protons and occurs for all projectile kinetic energies studied in this work, except for the highest one, $\approx$\,56.42 MeV.
For silicon projectiles that are initially highly ionized (Si$^{+12}$), RT-TDDFT noticeably overestimates SRIM and experiments;
in addition, there is no clear dependence on the lattice channel, and the magnitude of electronic stopping is even interchanged for two projectile kinetic energies ($\approx$2.79 MeV and $\approx$6.27 MeV).
For these, $\left<011\right>$ stopping is higher, while it is lowest for the other kinetic energies.
Conversely, for initially neutral Si projectiles, Fig.\ \ref{fig:Se_Si} shows that electronic stopping from RT-TDDFT is significantly lower and much closer to experiment and SRIM, resulting, however, in an overall underestimation.
Remarkably, the dependence of electronic stopping on the lattice channel discussed above for protons is restored in this case.
Finally, our results agree very well with those reported for low-kinetic energy silicon projectiles by Lim \emph{et al.}\ \cite{Lim_2016}, when the same simulation parameters are used.
However, since that work only accounts for the four valence electrons of silicon on the $n$=3 shell, the magnitude of electronic stopping is underestimated by more than 100\,\% compared to SRIM.
This implies that the $n$=2 shell of silicon target atoms interacts with projectiles with kinetic energies in the range shown in Fig.\ \ref{fig:Se_Si}.

Interestingly, Fig.\ \ref{fig:Se_Si} also shows that the dependence of electronic stopping on the initial projectile charge \emph{disappears} for off-channeling silicon projectiles.
In this case electronic stopping falls in between the results for initially neutral and highly ionized channeling projectiles and agrees more closely with experiment and SRIM.
As a consequence, we find that except for the highest kinetic energy, electronic stopping of initially charged heavy projectiles is \emph{significantly larger} on \emph{channeling} trajectories than on off-channeling trajectories.
While this is the opposite of what we discussed above for light (proton) projectiles, such an inverted behavior has indeed been observed for electronic stopping power of highly charged U ions channeling in Si \cite{Ray_2011}.
Below we explain this by developing a detailed understanding of the femto-second real-time dynamics of the projectile charge state, its dependence on the projectile trajectory, and its influence on electronic stopping.
We also note that RT-TDDFT for off-channeling projectiles still significantly underestimates the magnitude and projectile energy of the electronic stopping power maximum observed in experiment \cite{IAEA_Si_data} and SRIM \cite{Ziegler_2010}, see Fig.\ \ref{fig:Se_Si} at kinetic energies $\gtrapprox$ 6 MeV.
Investigating the origin of this disagreement in detail is the goal of a future study.

\subsection{\label{subsec:chargedyn}Femto-second dynamics of the projectile charge}

\begin{figure}[!hbt]
\includegraphics[width=0.9\columnwidth]{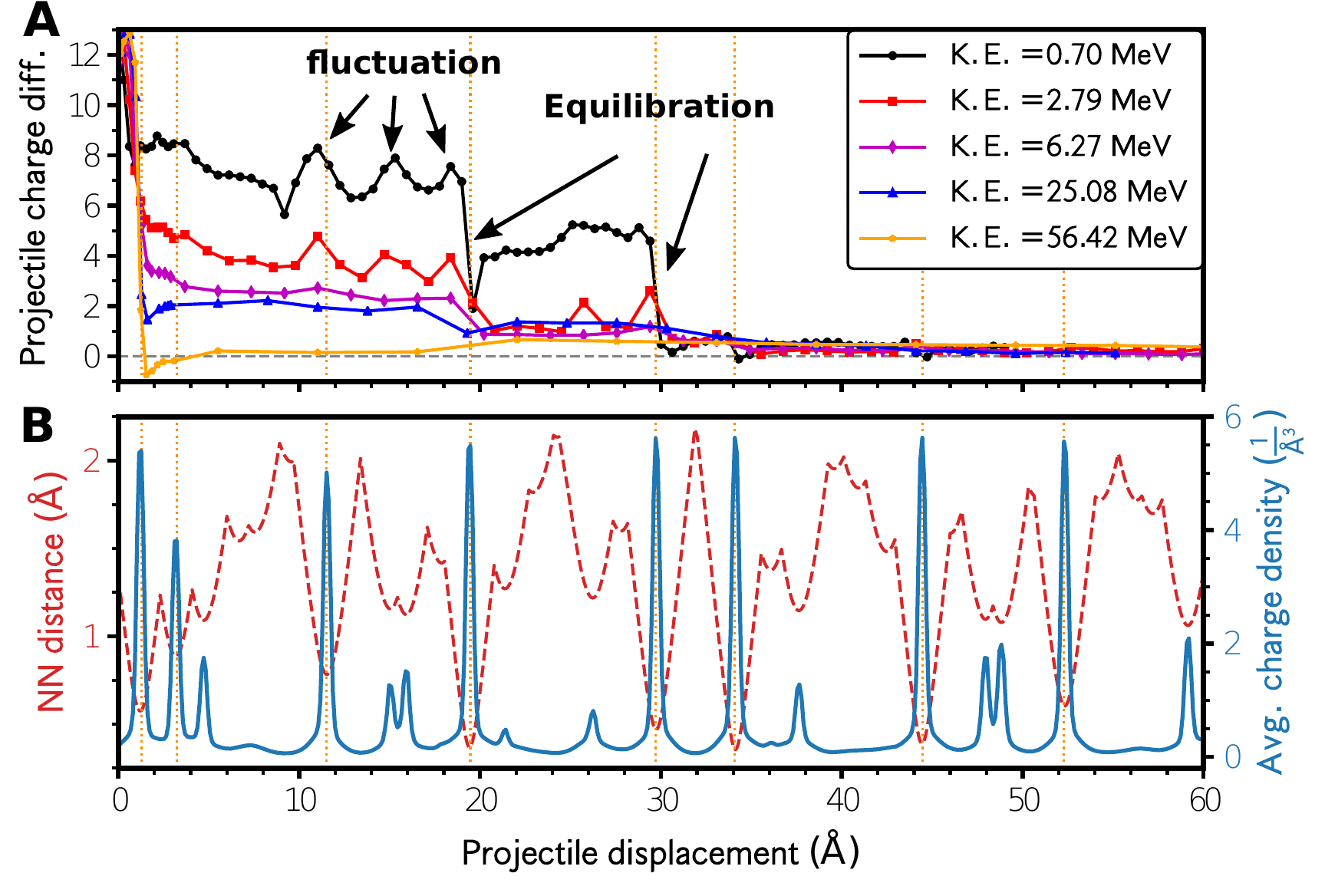}
\caption{\label{fig:traj_charge_1nn_rho}
Charge dynamics for self-irradiated Si.
(A) Position-dependent difference of charge state for initially highly ionized and neutral projectiles on an off-channeling trajectory.
Black circles, red squares, magenta diamonds, blue triangles, and orange stars indicate projectile kinetic energies (K.E.) of 0.70, 2.79, 6.27, 25.08, and 56.42 MeV respectively.
(B) First-nearest-neighbor (NN) distance (red dashed) and average charge density (blue solid) for the projectile on that same trajectory.
Strong peaks of average density for short NN distances coincide with significant changes in charge state, as highlighted by vertical orange dashed lines.
}
\end{figure}

As discussed for Fig.\ \ref{fig:Se_Si}, we find that the \emph{initial} projectile charge state strongly affects electronic stopping of channeling \emph{heavy} projectiles.
Using the Lindhard-Winther model, Eq.\ \eqref{eq:lindhard}, within which electronic stopping depends quadratically on the charge state of the projectile, this indicates that charge equilibration depends on the projectile trajectory.
In order to understand this surprising behavior, we analyze the femto-second real-time dynamics of the projectile charge in Fig.\ \ref{fig:traj_charge_1nn_rho}A.
This shows the time-dependent difference of the charge of initially highly ionized Si$^{+12}$ and initially neutral Si off-channeling projectiles, computed from time-dependent electron densities using the DDEC6 method \cite{Manz_2016}.
From this, extremely fast charge equilibration for the different projectile kinetic energies studied here becomes evident:
In all cases the charge states differ by $\approx$12.0 at first, but within only 1.55 fs (for the slowest projectile in Fig.\ \ref{fig:traj_charge_1nn_rho}A) or less, the initially highly ionized Si projectile acquires electrons and the initially neutral one loses electrons when traveling through the material (see Fig.\ S2  in Supplemental Material at [URL will be
 inserted by publisher] for separate dynamics depicted).

Fig.\  \ref{fig:traj_charge_1nn_rho}A also reveals important details of charge equilibration.
The first significant changes of the charge state appear for projectile penetration depths as small as 1.27 \AA.
After that, the charge state of initially neutral Si has approximately reached equilibrium already and subsequent dynamics is attributed mostly to initially highly charged Si (see Fig.\ S2 in Supplemental Material at [URL will be
 inserted by publisher] for separate dynamics of initially neutral and ionized Si projectile).
After a long quasi-equilibrium between 4 and 20 \AA, another series of equilibration processes appears around penetration depths of 20.0, 30.0, and 34.0 \AA.
These sudden equilibration events also occur at exactly the same projectile positions, independent of projectile velocity, and we conclude that charge equilibration is associated with penetration depth rather than time after impact.
Notably, faster projectiles are closer to charge equilibrium than slower projectiles early on (between 2 and 20 \AA\ in Fig.\ \ref{fig:traj_charge_1nn_rho}A), however, once the charge state differs by only +1 from equilibrium (around 30 \AA\ in Fig.\ \ref{fig:traj_charge_1nn_rho}A), the subsequent behavior becomes independent of the projectile velocity.
After vanishing around 34 \AA, the charge-state difference remains very small as the projectile travels further.

We explain the dependence on penetration depth and the occurrence of sudden equilibration events using the spatial distribution of the charge density of the target material.
To this end, Fig.\ \ref{fig:traj_charge_1nn_rho}B shows the distance of the projectile to first-nearest-neighbor atoms and the average charge density encountered by the projectile along the off-channeling trajectory.
Large changes of the charge state correlate with close spatial proximity to lattice atoms (see vertical orange dashed lines in Fig.\ \ref{fig:traj_charge_1nn_rho}B) and with the corresponding large local charge density attributed to highly localized semi-core electrons near target atoms.
Thus, from Fig.\ \ref{fig:traj_charge_1nn_rho} we conclude that charge equilibration of highly ionized Si$^{+12}$ requires close spatial proximity of the projectile to target ions and that equilibration by attracting electrons is mediated by semi-core electrons of the target.
This argument is further supported by the observation that the separation between projectile and target atom needs to fall below a certain value, i.e., the charge density the projectile interacts with needs to exceed a certain value, for equilibration to happen.
This is evidenced by a slightly lower electron density peak near 11.8 \AA\ that merely triggers fluctuations of the subsequent charge dynamics (indicated by the arrows in Fig.\ \ref{fig:traj_charge_1nn_rho}A) for two kinetic energies, but no actual equilibration event.
To provide further proof, we compare to simulations using pseudopotentials with only four valence electrons per Si atom of the target material (see Fig.\ S3 in Supplemental Material at [URL will be inserted by publisher] for explicit comparison) and find that the charge state of initially highly ionized Si$^{+12}$ indeed remains much higher due to the exclusion of charge transfer from pseudized core electrons of the target material.

Similar analysis of the charge state for \emph{channeling} Si projectiles shows that a significant reduction of the projectile-charge difference also occurs early on---within the first 2.0 \AA.
This is only slightly deeper than for off-channeling projectiles and we find again that the charge state after this first drop depends on projectile kinetic energy.
As discussed for off-channeling projectiles, the position of this first drop is independent of projectile kinetic energy (see Fig.\ S4 in Supplemental Material at [URL will be inserted by publisher] for charge state dynamics of different channeling silicon projectiles with different kinetic energies).
Our analysis of the subsequent dynamics shows further, rather constant equilibration up to about 40 \AA\ for projectiles on $\left<001\right>$ and $\left<111\right>$ channels.
Contrary to this, there is very little equilibration after the initial drop for a projectile on a $\left<011\right>$ channel, for which the electron density along the projectile path is smallest (see Fig.\ S1A in Supplemental Material at [URL will be inserted by publisher] for explicit data).
This further corroborates our interpretation that charge equilibration is connected to the electron density the projectile interacts with.
Interestingly, contrary to off-channeling projectiles, our analysis reveals that the projectile charge state remains fairly constant after 40 \AA\ for \emph{all} channels.
This leads to a different equilibrium charge state of channeling and off-channeling projectiles for most projectile kinetic energies.
We explain this difference by a lack of interactions of channeling projectiles with core electrons, since they never approach target ions closely enough and, thus, only equilibrate via interactions with valence electrons of the target.

\subsection{\label{subsec:chargeeq}Equilibrium projectile charge and electronic stopping}

\begin{figure}[!hbt]
\includegraphics[width=0.9\columnwidth]{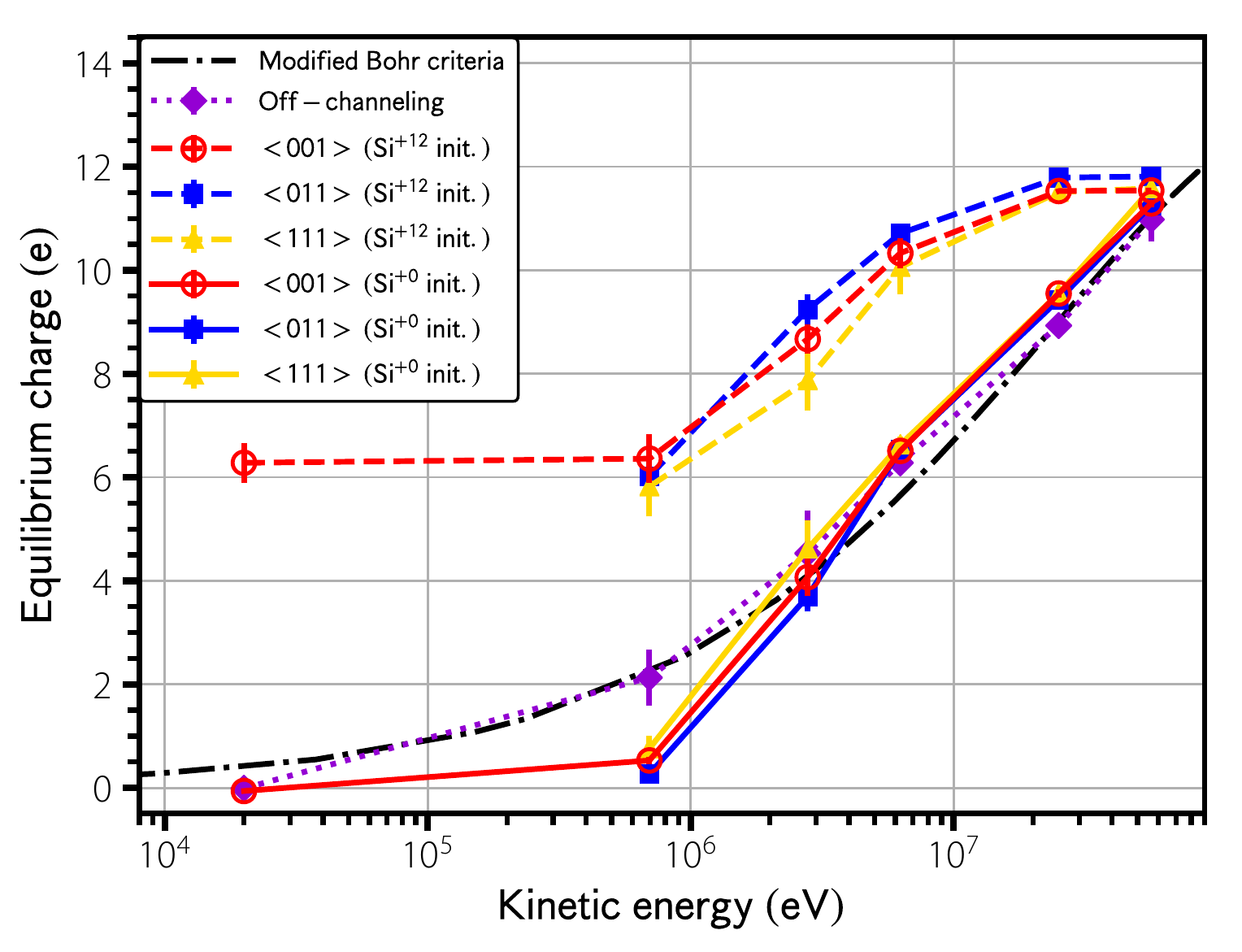}
\caption{\label{fig:KE_CHG_Si_Si}
Equilibrium projectile charge state of initially highly ionized (dashed lines) and neutral (solid lines) projectiles for $\left<001\right>$ (red circles), $\left<001\right>$ (blue squares), $\left<111\right>$ (gold triangles), and off-channeling (violet diamonds) trajectories.
For off-channeling projectiles (dotted line) the equilibrium value is independent of the initial charge state.
We also compare to an analytical result based on Bohr's stripping criterion (black dot-dashed line, see text for details).
}
\end{figure}

While the above analysis showed that initially ionized Si$^{+12}$ projectiles attract electrons and initially neutral projectiles lose electrons as they travel through the target, neither the emerging dynamic nor the final equilibrium charge are known \emph{a priori} when materials are irradiated in experiment.
This has been a longstanding problem, since electronic stopping, for instance within Lindhard-Winther theory \cite{Lindhard_1964}, Eq.\ \eqref{eq:lindhard}, explicitly depends on the projectile charge.
However, in practice, it is only described using semi-empirical models \cite{Sigmund_2014}.
In the following, we use our accurate first-principles results to provide insight into how the equilibrium charge emerges from the dynamic charge and
we disentangle the underlying connection between projectile charge, semi-core electron contributions, and electronic stopping.

To this end, we first compute the \emph{equilibrium} projectile charge by averaging the dynamic charge over the same spatial range used for computing electronic stopping;
this ensures a fair comparison with electronic-stopping results.
Fig. \ref{fig:KE_CHG_Si_Si} illustrates that slow initially-ionized projectiles attract more electrons from the target material and become less ionized in equilibrium, but remain more ionized for higher projectile kinetic energies.
For the highest kinetic energy of 56.42 MeV studied here (see Fig.\ \ref{fig:KE_CHG_Si_Si}), the projectile stays entirely ionized (except for the 1$s$ shell), i.e., Si$^{+12}$.
Contrary, initially-neutral projectiles lose electrons through interactions with the electronic system of the target and remain neutral or weakly charged if they are slow.
The fastest projectiles studied in this work lose all electrons and become entirely ionized, except for the 1$s$ shell (see Fig.\ \ref{fig:KE_CHG_Si_Si}), i.e., Si$^{+12}$.

Fig.\ \ref{fig:KE_CHG_Si_Si} also illustrates a dependence on the projectile trajectory:
Slow Si$^{+12}$ projectiles attract significantly fewer electrons
when they move on a channeling compared to an off-channeling trajectory.
This difference becomes smaller for faster projectiles, since all fast projectiles are overall more ionized.
We also find consistently smaller equilibrium charge for Si$^{+12}$ on a $\left<111\right>$ channel along which the average electron density is large, compared to highest equilibrium charge for a $\left<011\right>$ channel with small average electron density.
Interestingly, our results show a very different trend for initially neutral channeling projectiles:
There is no clear dependence of the equilibrium charge state on the specific lattice channel.
Even more strikingly, for \emph{off-channeling} projectiles, the equilibrium charge becomes completely independent of the initial charge state across the entire kinetic-energy range studied here and 
is very close to that of initially neutral, channeling projectiles, except for one data point at a kinetic energy of 0.7 MeV.
This implies that highly ionized Si$^{+12}$ projectiles equilibrate their charge state by acquiring \emph{semi-core} electrons from the target, which they can only interact with when approaching target atoms closely, e.g., on an off-channeling trajectory.
The explicit involvement of semi-core electrons of the target material is further supported by Fig.\ S3 (see Supplemental Material at [URL will be inserted by publisher] for explicit comparison), which rules out effects arising from the mere proximity of projectile and target ion. 
Our results also imply that stripping electrons off the projectile only requires some electron density to scatter, but whether the scattering involves explicit semi-core electrons of the target is not important.
This is also supported by Fig.\ S3 which shows that the equilibrium charge state of initially neutral projectiles does not depend on the number of valence electrons used to describe the target material.

Furthermore, we find excellent \emph{quantitative} agreement between our results for stripping off electrons from off-channeling and initially neutral channeling projectiles and Bohr's stripping model.
This is an analytical model that predicts the velocity-dependent equilibrium charge of the projectile upon interaction with a target material \cite{Bohr_1948}.
It relies on a hydrogenic model and approximates the outermost shell principal quantum number as cubic root of the atomic number $Z$ of the projectile.
In this model, electrons with orbital velocities smaller than the projectile velocity are stripped off and in its modified analytic form \cite{Bohr_1948,Sigmund_2014} it reads
\begin{equation}
\label{eqn:bohr_stripping}
Z_\mathrm{eff}(v) \approx Z\left(1-\exp{\left[-\frac{v}{Z^{2/3}v_{0}}\right]}\right).
\end{equation}
Here, $Z_\mathrm{eff}$, $Z$, and $v$ are the equilibrium/effective charge state, atomic number, and velocity (in atomic units) of the projectile, respectively, and $v_{0}$ is the Bohr velocity of the hydrogen atom \cite{Bohr_1948,Pierce_1968, Sigmund_2014}.
This model qualitatively explains the higher equilibrium charge state of faster projectiles without any parameters describing the target material.

Interestingly, our results also show that the equilibrium charge for an initially highly charged projectile that \emph{attracts} electrons coincides with that of a projectile that loses electrons and, thus, is also successfully described by Bohr's criterion.
Fig.\ \ref{fig:KE_CHG_Si_Si} shows that this is only the case for projectiles that can actually reach their equilibrium charge state by attracting semi-core electrons, e.g., when traveling on an off-channeling trajectory.
These results are in agreement with experimental results \cite{Martin_1969} for O$^{+8}$ ion projectiles in Si target material, showing that initially highly ionized and neutral off-channeling projectiles reach the same charge state.
This experiment also shows that an initially highly ionized channeling ion has a higher equilibrium charge state than an off-channeling one and that an initially weakly charged channeling ion reaches the same equilibrium charge state as the off-channeling one, again confirming our results.

\begin{figure}
\includegraphics[width=0.99\columnwidth]{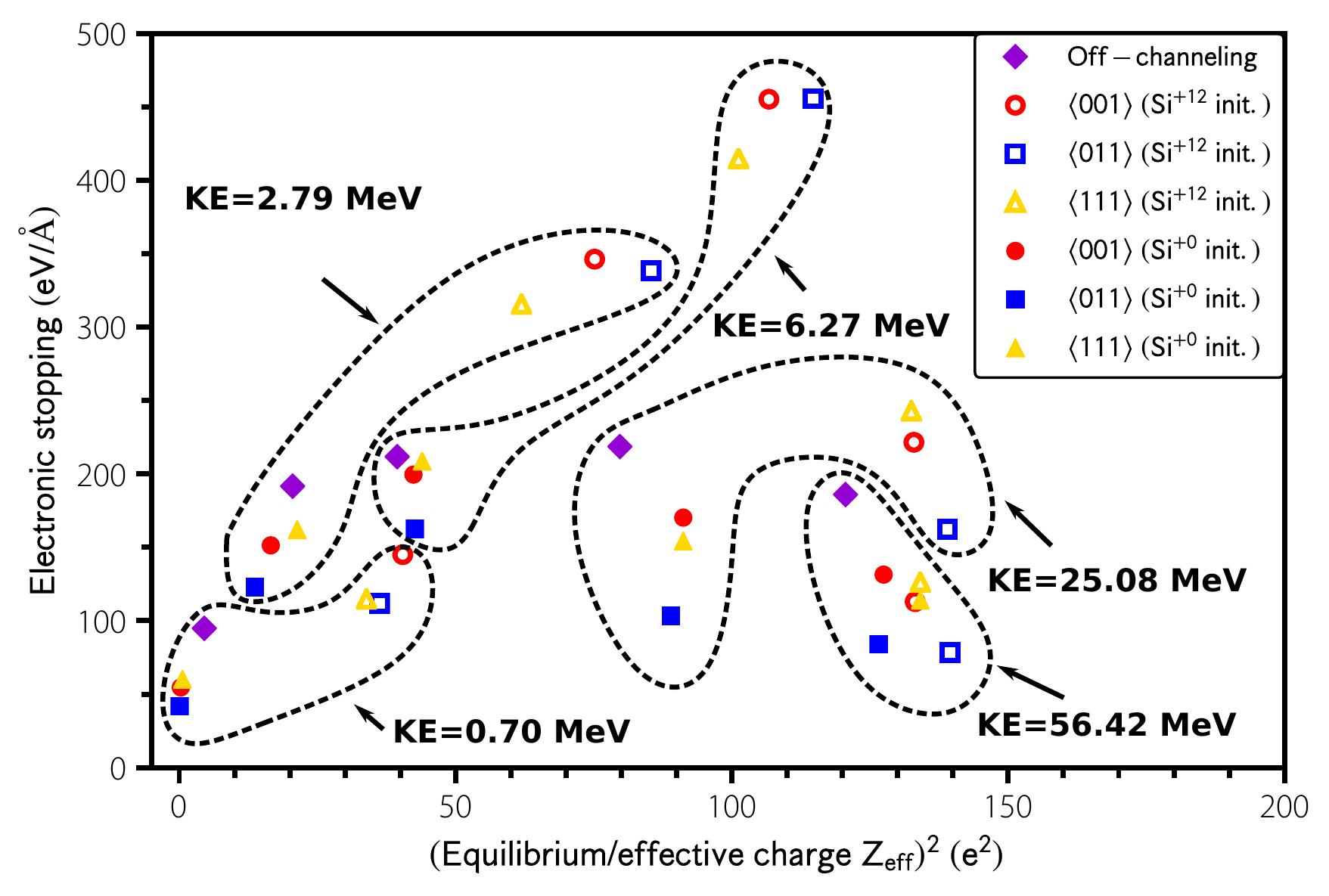}
\caption{\label{fig:SP_Chg2}
Electronic stopping versus squared equilibrium effective projectile charge for self-irradiated Si.
Black dashed lines group the data by projectile kinetic energy.
Violet filled diamonds indicate off-channeling Si projectiles.
Red circles, blue squares, and gold triangles indicate $\left<001\right>$, $\left<011\right>$, and $\left<111\right>$ channeling projectiles, respectively.
Filled and open symbols indicate the initial charge state of the projectile.
}
\end{figure}

Finally, our first-principles results for \emph{equilibrium} charge states provide deeper understanding of electronic stopping:
Within Lindhard-Winther theory \cite{Lindhard_1964}, Eq.\ \eqref{eq:lindhard}, electronic stopping scales linearly with squared projectile charge $Z_\mathrm{eff}^{2}$ and linearly with $n\cdot L(n)$, for a given projectile kinetic energy.
In practice, the uniform electron gas model with \emph{average} charge density of valence electrons of the whole system is used for $n$ \cite{Correa_2018}.
However, for projectiles with high kinetic energy, it is necessary to account also for deeper core electrons.
Similarly, the projectile experiences different local charge density on different trajectories, giving rise to different stopping \cite{Ullah_2015, Lee_2018}.
Unfortunately, there is no \emph{a priori} knowledge on how to choose $n$, since there is no definition of the effective charge density $n$ a projectile interacts with when moving through an inhomogeneous charge density distribution, as is the case for a Si projectile in a Si target material.
Furthermore, the effective projectile charge $Z_\mathrm{eff}$ is velocity-dependent and only  equivalent to $Z$ for the fully ionized case at very high projectile kinetic energies.

Hence, to connect with Lindhard-Winther theory \cite{Lindhard_1964}, we investigate the relation between electronic stopping and $Z_\mathrm{eff}^{2}$ for all trajectories of Si projectiles and different initial charge states in Fig.\ \ref{fig:SP_Chg2}.
We note that the slopes that can be assigned to each group of data points of the same kinetic energy (indicated by dashed lines in Fig.\ \ref{fig:SP_Chg2}) represent the $n\cdot L(n)$ term in Eq.\ \eqref{eq:lindhard}.
Though the effective electron density $n$, experienced by the Si projectile under the various conditions, is unknown, we distinguish three different kinetic energy regimes:
For low kinetic energy (KE $\lessapprox$ 6.27 MeV), Fig.\ \ref{fig:SP_Chg2} clearly shows that electronic stopping linearly depends on $Z_\mathrm{eff}^{2}$, i.e., $n\cdot L(n)\approx\text{const.}$\ for different trajectories, suggesting that differences in $Z_\mathrm{eff}$ significantly affect electronic stopping.
On the other hand, for high kinetic energies around 56.42 MeV, Si projectiles are mostly ionized, i.e., have similar equilibrium charge independent of their trajectory.
In this case, differences in electronic stopping are dominated by the effective charge density $n$ the projectile interacts with.
This is similar to electronic stopping of light ions and, accordingly, off-channeling Si projectiles in this kinetic-energy range experience the largest stopping, while those on $\left<011\right>$ channels experience the lowest.
For intermediate kinetic energies, we find a balance of both contributions:
Similar to the case of low kinetic energy, initially highly ionized \emph{channeling} Si ions have larger electronic stopping than initially neutral ones, within the same lattice channel.
However, in contrast to the case of low kinetic energy, initially highly ionized Si projectiles on a $\left<011\right>$ channel experience lower electronic stopping than initially neutral ones on $\left<001\right>$ and $\left<111\right>$ channels, despite their larger equilibrium charge.
Similarly, off-channeling projectiles experience larger stopping than most channeling ones, despite showing the lowest charge state.
This suggests that, at intermediate kinetic energies, neither effective charge nor effective electron density alone dominates electronic stopping.
Finally, while the charge state for initially neutral channeling and off-channeling projectiles is approximately the same, electronic stopping is larger for off-channeling projectiles  across the entire kinetic-energy range.
We attribute this to contributions from the core electrons of the target, which require spatial proximity of projectile and semi-core electrons, as discussed before \cite{Schleife_2015, Yost_2017}.

\subsection{\label{ssxn:md-results}Defect dynamics in single ion-strike damage events}

\begin{figure*}[!hbt]
  \centering
  {\includegraphics[width=0.6\textwidth]{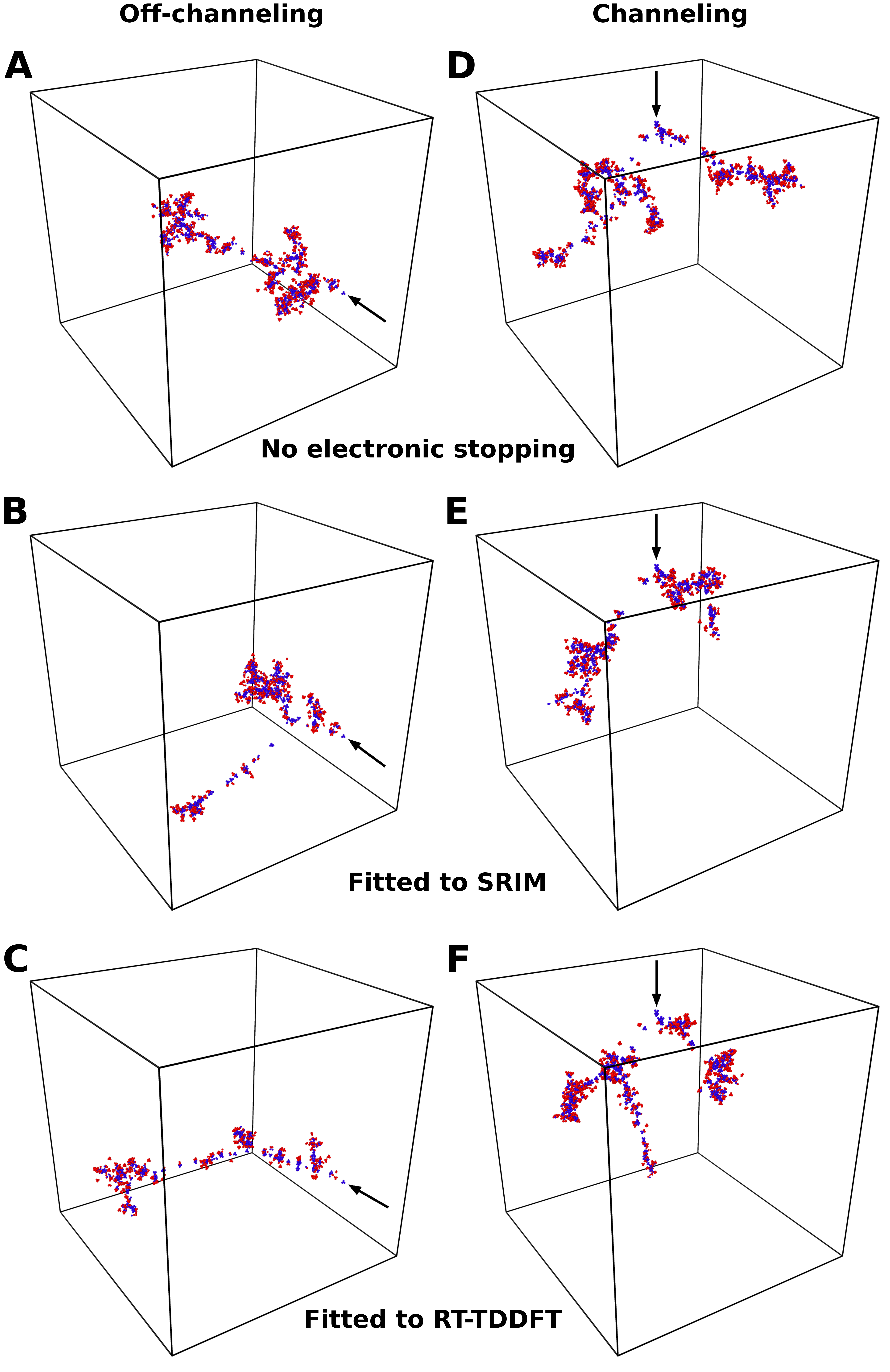}}								\quad
\caption{\label{fig:defect-shape}
Representative displacement cascade structures and average total number of defects for off-channeling (A, B, C) and $\langle 001 \rangle$ channeling directions (E, F, G) for a projectile with an initial kinetic energy of 20 keV.
Electronic stopping is neglected (A, D), fitted to SRIM (B, E), and fitted to RT-TDDFT for the neutral charge state (C, F).
Coloring outlines the cascade structure where blue indicates a local vacancy, i.e., atoms that have lost a nearest neighbor (coordination number $< 4$), red indicates a local self-interstitial, i.e., atoms that have gained a nearest neighbor (coordination number $> 4$). Arrows indicate the initial PKA direction.
}
\end{figure*}

Having established the role of core electrons, projectile charge state, and projectile trajectory on electronic stopping power, we now examine how these affect development and evolution of displacement cascades.
We explore the consequences of energy deposition into the electronic system of the target during the earliest stages of radiation damage on the development of an initial radiation damage event across multiple length and time scales.
As the initial radiation damage event develops, primary knock-on atom (PKA) and subsequent knock-on events lead to displaced atoms and the creation and evolution of complex cascade structures.
These lead to creation and evolution of point defects, dislocation loops, defect clusters, or voids, but are 
critically affected by the initial energy deposition.
In addition, the final state of this cascade is extremely important because it is the starting point for subsequent defect diffusion, agglomeration, and annihilation that form the basis of observable effects of radiation in materials \cite{Krasheninnikov_2010, Averback_1997,Darkins_2018}.
Here we incorporate electronic stopping from SRIM and limiting cases from RT-TDDFT into molecular dynamics (MD) simulations to study the implications of how electronic stopping is represented for defect dynamics in silicon, and to compare to the standard approach of neglecting electronic stopping entirely.
In this work, electronic stopping is incorporated into MD using the inelastic energy loss (IEL) approach;
test calculations using a two-temperature model have shown that the differences compared to IEL are small in the low recoil-energy regime studied here.

The electronic stopping results for a slow Si projectile traveling through bulk Si in Fig.~\ref{fig:Se_Si} suggest capturing electronic stopping in the IEL approach using a proportionality constant $\gamma_e$.
Here we assume $\gamma_e$ to be a linear function of the atomic velocity of the projectile,
\begin{equation}
\label{eqn:stopping-fit}
\gamma_{e} = a v_i + b.
\end{equation}
Here, $a$ and $b$ are fitting parameters, and the total electronic stopping power, $\gamma_{e} v_i$, is a quadratic function of the absolute value of the atomic velocity $v_i$, where $i$ indexes atoms.
Using the condition that $\textbf{S}_{e}(v_{i} = 0)=0$, we fit to three limiting cases, i.e., (i) an off-channeling projectile for which we found charge equilibration, (ii) an initially neutral, and (iii) an initially Si$^{+12}$ projectile, both on a $\langle001\rangle$ trajectory.
Fits to our RT-TDDFT data yield noticeably different values of $a=0.0$ eV$\cdot$ps$^{2}$/\AA$^{3}$, $b=4.9\times 10^{-3}$ eV$\cdot$ps/\AA$^{2}$ for off-channeling projectiles, $a=5.0\times 10^{-8}$ eV$\cdot$ps$^{2}$/\AA$^{3}$, $b=1.4\times 10^{-3}$ eV$\cdot$ps/\AA$^{2}$ for $\langle001\rangle$ channeling ${\rm{Si}}^{+0}$, and $a=7.0\times 10^{-8}$ eV$\cdot$ps$^{2}$/\AA$^{3}$, $b=5.1\times 10^{-3}$ eV$\cdot$ps/\AA$^{2}$ for $\langle001\rangle$ channeling ${\rm{Si}}^{+12}$ projectiles.
A fit to SRIM yields $a=4.0\times 10^{-8}$ eV$\cdot$ps$^{2}$/\AA$^{3}$, $b=3.3\times 10^{-3}$ eV$\cdot$ps/\AA$^{2}$.
All previous IEL studies assumed a constant $\gamma_{e}$, here instead we provide a first-order correction through a velocity-dependent $\gamma_{e}$ in Eq. (6).  
While this correction is minor for low kinetic-energy range (see Fig. 3 and as in the case for our MD simulations with a projectile with an initial kinetic energy of 20 keV), this correction becomes increasingly important as the recoil energy increases (see Fig. S6  in the Supplemental Material at [URL will be inserted by publisher] for different fitted functions), yielding significant differences in defect production.
We then use the fitting results for $a$ and $b$ to perform MD simulations with otherwise identical simulation conditions, i.e., identical initial velocities in terms of thermal noise and identical direction and magnitude of the PKA.
Results discussed hereafter, thus, describe differences in the stages of the cascade developments \emph{solely} due to the underlying electronic-stopping physics.
For the off-channeling case, the same random atom is given a 20 keV recoil energy and a direction towards the center of the domain.
For the channeling direction, a different atom is chosen from that of the off-channeling case, but we chose the same atom and energy for all the channeling simulations.

The resulting representative cascade structures and the displacement cascade damage are shown in Fig.\ \ref{fig:defect-shape}.
We analyze the spatial distribution of defects constituting the final cascade by comparing PKA MD simulations without electronic stopping and with electronic stopping from SRIM and RT-TDDFT for off-channeling and $\langle 001 \rangle$ channeling directions.
These MD simulations were performed using otherwise identical initial conditions, e.g., velocity distribution, selected PKA energy, and projectile direction, and, thus, any difference in cascade structure is solely attributed to the representation of electronic stopping.
It is interesting to note that, in all cases, the atoms indicating local vacancies are concentrated at the core of the cascade while the atoms indicating local self-interstitials envelop these regions.
In all of the simulations performed, multiple sub-cascade branches can be visually identified as comprising the full cascade structure and are also seemingly aligned along specific crystallographic directions within the bulk crystal or contained within small amorphous pockets (see Fig.\ \ref{fig:defect-shape}).

To quantify the overall principal directions and relative shapes of these combined sub-cascades, we performed a principal component analysis (PCA) \cite{Abdi_2010, Jolliffe_2014} using the positions of the atoms identified to be constituting the primary cascade structure.
This PCA transforms the atomic positions of the discrete defects in the cascade to lie along three principal direction vectors (i.e., lines of best fit) such that the variance of the atomic positions along these lines is minimized.
The cascade structure can be approximated as an ellipsoid with its three principal axes $A$, $B$, and $C$ given by PCA.
Relative geometric shapes of the cascades are then characterized via the aspect ratios $A/B$, $A/C$, and $B/C$.
Cascade size is described by their volume, computed by multiplying the average atomic volume by the number of displaced atoms.
Explicit results of this volume calculation, ellipsoid aspect ratios, and $1^{\mathrm{st}}$ PCA vector describing the primary orientation of the ellipsoid are summarized in table S1.

We find that the underlying electronic-stopping physics and projectile trajectory affect cascade size, shape, and orientation in MD simulations:
For off-channeling projectiles, the cascades formed using electronic stopping fitted to RT-TDDFT are consistently more compact (i.e., they have the smallest volume) than those formed when electronic stopping is fitted to SRIM or neglected entirely.
When electronic stopping is accounted for, this mechanism dissipates part of the projectile kinetic energy and, thus, leads to less extensive lattice defects. 
Additionally, in the case of off-channeling projectiles the general shape of the cascades as described using aspect ratios resembles a flattened prolate ellipsoid when electronic stopping is neglected.
Contrary, for channeling projectiles with electronic stopping fitted to RT-TDDFT data for the highly ionized Si$^{+12}$ projectile, the aspect ratios suggest an oblate ellipsoid.
Finally, the incorporation of electronic stopping can also have a significant impact on the orientation of the resulting cascade.

\begin{figure*}[!bht]
\centering
{\includegraphics[width=0.9\textwidth]{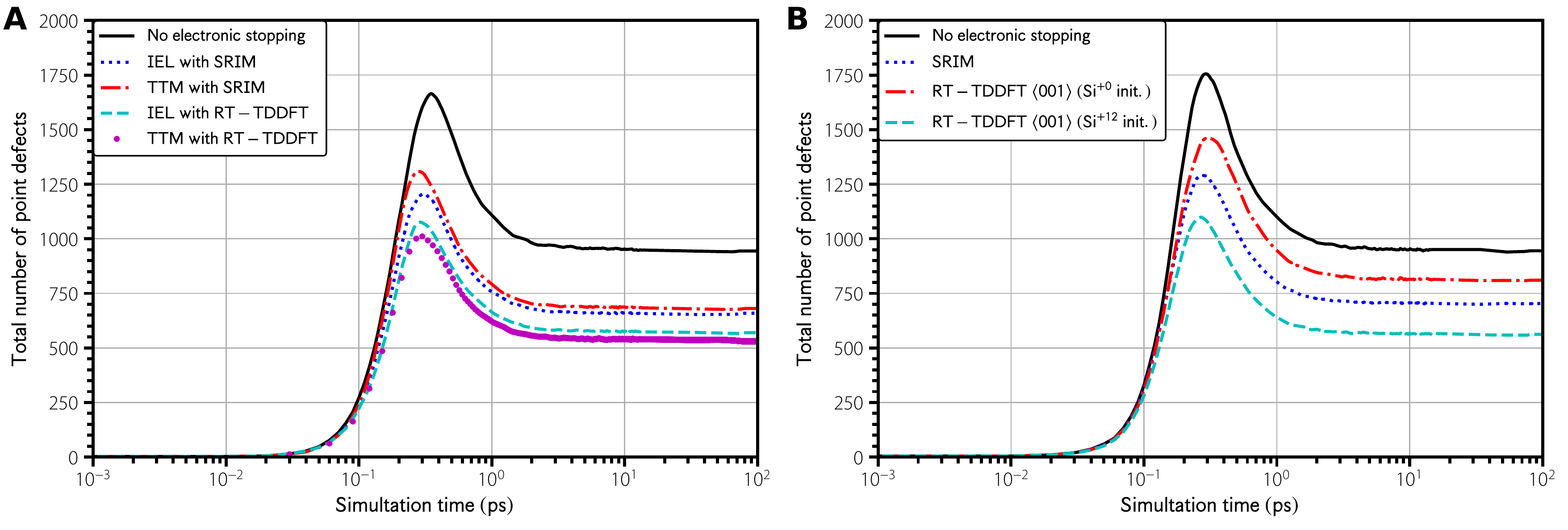}}
\caption{\label{fig:defect-count}
Average total number of defects as a function of time for off-channeling (A) and channeling (B) trajectories. Electronic stopping is neglected (black solid lines), fitted to SRIM (blue short dashed and red dot-dashed lines), and fitted to RT-TDDFT (cyan long dashed line and violet circle dots).
}
\end{figure*}

Aside from analyzing the cascade structure, we characterize the total number of Frenkel pairs produced by the PKA by defect count, displacement cascade size, and associated shape \cite{Stewart_2018}.
This shows (see Fig.\ \ref{fig:defect-shape}) that after the cascade is initiated, the number of defects rapidly increases to the peak damage regime within $\sim 3$\,--\,4 ps.
Immediately after this peak damage regime, the energy of the atomic system continues to dissipate throughout the surrounding bulk region and, due to recombination events, the total number of defects subsequently decreases and stabilizes for the remainder of the simulation.
Differences in the defect count presented in Fig.\ \ref{fig:defect-count} indirectly indicate that for both off-channeling and channeling conditions the representation of the electronic stopping has a direct effect on the partitioning of the energy transferred to the lattice as the cascade develops.

In particular, our findings show that capturing the appropriate electronic stopping physics within MD simulations, e.g., via the parameter $\gamma_{e}$, is important to describe the defect cascade.
The resulting average total number of point defects in the final cascade structure illustrates quantitative differences, depending on channeling vs.\ off-channeling trajectory in MD, projectile charge state, and how electronic stopping power is described, i.e., using fits to SRIM vs.\ RT-TDDFT (see table\ S1 of the Supplemental Material at [URL will be
 inserted by publisher] for detailed numerical results).
We find that the number of defects generated at peak damage is greatly reduced when incorporating electronic stopping effects (see Fig.\ \ref{fig:defect-count}).
Neglecting electronic stopping we found on average $\sim 944$ point defects for both the off-channeling and channeling simulations.
However, when accounting for electronic stopping effects, the resulting average number of point defects is between 530 and 680 for off-channeling and between 560 and 810 for $\langle 001 \rangle$ channeling.
This is because for the off-channeling case, more electronic stopping occurs and, as such, less energy is deposited into the lattice to displace atoms, leading to fewer defects.
For the channeling case, electronic stopping is smaller than for off-channeling, but larger compared to the simulations without any electronic stopping.
Hence, we observe a moderate amount of defects created.
In addition, our data for $\langle 001 \rangle$ channeling shows large differences in the resulting maximum as well as equilibrium number of defects, depending on the two different equilibrium projectile charge states.
As discussed above, these two equilibrium charge states arise for initially highly charged and neutral channeling projectiles due to the lack of interactions with core electrons and, consequently, lead to two different values of electronic stopping.

These results are also consistent with the standard analytical approximation of damage production given by the model from Norgett, Robinson, and Torrens (NRT) \cite{Norgett_1975}, which is an important validation of the MD potential and implementation of electronic stopping.
This model is an accepted standard for estimating damage and it quantifies point defect count in a bulk crystal based on energy deposited into the system using a spherical cascade approximation.
Within the NRT model, the number of Frenkel pairs produced by an incident energetic particle is a monotonically increasing function of the recoil energy of the PKA, $E_{\mathrm{PKA}}$, given by
\begin{equation}
\label{eqn:nrt}
N_{\mathrm{NRT}} = 0.8 \left( E_{\mathrm{PKA}} - Q_{e^-} \right)/2E_{\mathrm{d}},
\end{equation}
where $Q_{e^-}$ is the total energy loss due to electronic stopping and $E_{\mathrm{d}}$ is the threshold displacement energy, approximated here as $E_{\mathrm{d}}=16.88$  eV for bulk Si using the Tersoff interatomic potential \cite{Miller_1994}.
Values for $Q_{e^-}$ and for NRT model predictions are tabulated in detail in table\ S1 of the Supplemental Material at [URL will be
 inserted by publisher].
This shows that accounting for energy dissipation by electronic stopping reduces the effective energy deposited in system and, thus, the defect count will be lower according to Eq.\ \eqref{eqn:nrt}.

Finally, the data in Fig.\ \ref{fig:defect-count} shows that while including or neglecting electronic stopping in the MD description strongly affects the cascade, there is also a notable difference depending on whether SRIM or RT-TDDFT is used to parametrize electronic stopping.
Results from MD calculations with electronic stopping fitted to SRIM, which contains no crystal structure or charge-state information, fall in between those of the different charge states, as shown in Fig.\ \ref{fig:defect-count}B.
This underlines the need for precise models, such as RT-TDDFT, for dynamics of projectile charge and the resulting electronic stopping power.
Furthermore, for the off-channeling case in Fig.\ \ref{fig:defect-count}A, we compared the inelastic energy loss (IEL) approach to the two-temperature model (TTM) and found them within one standard deviation of each other.
This implies that within the linear stopping regime, there is no statistically meaningful difference between these two models in regards to the generation of displacement cascade defects.

\section{\label{sxn:discuss}Discussion}

Our multi-scale results, combining RT-TDDFT to predict electronic stopping and the inelastic energy loss method to incorporate it into full cascade molecular dynamics simulations, lead to a detailed picture of how core electrons participate in these processes and we find that they play a two-fold role:

First, they can directly absorb energy from fast projectiles and we confirm earlier studies for semiconductor and metal targets that showed this leading to (i) off-channeling stopping being largest at a given velocity and (ii) RT-TDDFT results for channeling projectiles underestimating SRIM at high kinetic energies \cite{Schleife_2015, Yost_2017,Ojanpera_2014,Ullah_2018}.
This is because only projectiles with high kinetic energy can excite strongly bound semi-core electrons and unlike valence electrons, semi-core electrons are localized near nuclei, so that only off-channeling projectiles approach them closely enough to interact with them.
Such an involvement of projectile core electrons in electronic stopping was also discussed for heavy projectiles before \cite{Ojanpera_2014,Ullah_2018}.

Second, our results explicitly show that core electrons are also critically important for equilibration of the projectile charge state of initially highly charged ions, depending on projectile trajectory, kinetic energy, and initial projectile charge state;
this subsequently affects electronic stopping.
For the projectile kinetic energies studied in this work, charge equilibration is fast ($\approx$ 1 fs) compared to any appreciable reduction in kinetic energy of the projectile ($<$ 0.5 \% on the same time scale). 
Hence, we discuss the equilibrium charge state of a projectile at a given kinetic energy, which intuitively should be independent of its initial charge, as well as its velocity dependence.
However, our results illustrate that this equilibrium emerges as an intricate balance between attracting and stripping off electrons.
Whether the projectile can reach equilibrium, thus, depends on projectile kinetic energy and trajectory.
Since we cannot explicitly distinguish between individual events of attracting and stripping off electrons in our RT-TDDFT simulations, we interpret the kinetic-energy dependence of stripping off electrons using the binding energy of electronic states in silicon.
Fig.\ \ref{fig:KE_CHG_Si_Si} shows that two ($3p$), four ($3s$+$3p$), ten ($2p$+$3s$+$3p$), and twelve ($2s$+$2p$+$3s$+$3p$) valence electrons of the Si projectile are completely stripped off from the initially neutral projectile for kinetic energies larger than about 1.4, 2.8, 25.1, and 56.4 MeV, respectively.
These results are consistent with threshold energies of 0.04 MeV ($3p$), 0.3 MeV ($3s$), 24.5 MeV ($2p$) and 42.1 MeV ($2s$), computed using Eq.\ \eqref{eqn:vth} and the corresponding atomic ionization energies of the Si projectile of 16.346, 45.142, 401.38 and 523.415 eV/atom, respectively \cite{NIST_2018}.
Due to band-structure and hybridization effects, this estimate based on atomic ionization energies is slightly worse for $3s$ and $3p$ valence electrons.

Our simulations also shed light on the dynamics of attracting electrons from the target:
Fig.\ \ref{fig:KE_CHG_Si_Si} shows that initially neutral projectiles with the lowest-kinetic energies are not stripped off their electrons.
Hence, the two different equilibrium charge states observed for initially charged and initially neutral channeling projectiles imply that in our simulations, valence electrons attracted from the target by slow, initially charged projectiles do not subsequently relax into projectile core states.
Our simulations might be too short to explicitly capture these effects:
The longest trajectory ends at about $\lessapprox$ 1 fs and, for instance, the auto-ionization rate is reported \cite{IAEA_ai_Si} as less than 0.4 fs$^{-1}$.
In addition, it is also reported in the literature that relaxation mechanisms, such as auto-ionization, require inclusion of memory effects in exchange and correlation \cite{Kapoor_2016} not captured by the adiabatic local-density approximation used here.
Similarly, we cannot distinguish whether the higher equilibrium charge of initially ionized silicon projectiles with high kinetic energies (see Fig.\ \ref{fig:KE_CHG_Si_Si}) is due to a smaller capture cross section, or subsequent stripping off of electrons.

We also note that we exclude the possibility that Si\,1$s$ core states contribute.
Treating a projectile of mass $m$ that travels through a periodic lattice with spatial periodicity of $\lambda$=1.34 \AA\ for $\left<001\right>$ channeling as a time-dependent perturbation to the target material \cite{Lim_2016}, allows computing the threshold velocity for excitations of electrons across a certain energy gap $\Delta$,
\begin{equation}
  KE=\frac{1}{2}m v^{2}=\frac{1}{2}m\left(\frac{\lambda \Delta}{h}\right)^2,
\label{eqn:vth}
\end{equation}
where $h$ is Planck's constant. 
We estimate that due to their large binding energy of $\Delta\approx$ 2.5 keV \cite{NIST_2018}, Si\,1$s$ electrons only contribute to electronic stopping of silicon projectiles with kinetic energies of about 0.9 GeV or higher.
Similarly, Si\,1$s$ electrons only contribute to electronic stopping of protons with kinetic energies of about 32.7 MeV or higher.

Finally, we note that we verified that initially neutral and initially ionized hydrogen projectiles in silicon equilibrate to the same charge state for channeling and off-channeling trajectories (see Fig.\ S5B in Supplemental Material at [URL will be inserted by publisher] for equilibrium charge states of protons with different kinetic energies).
We explain this with the low binding energy of the H\,$1s$ electron, which is comparable to binding energies of valence electrons of silicon.
Hence, even \emph{fully} ionized light projectiles have no deep core states to fill and, thus, fully equilibrate through interactions with valence electrons of the target, not requiring involvement of core states.
This is consistent with previous studies that reported no difference in electronic stopping of neutral hydrogen atoms vs.\ protons \cite{Schleife_2015} and confirms the fundamentally different charge equilibration and electronic stopping of light vs.\ heavy projectiles.

\begin{figure}[hbt!]
{\includegraphics[width=0.99\columnwidth]{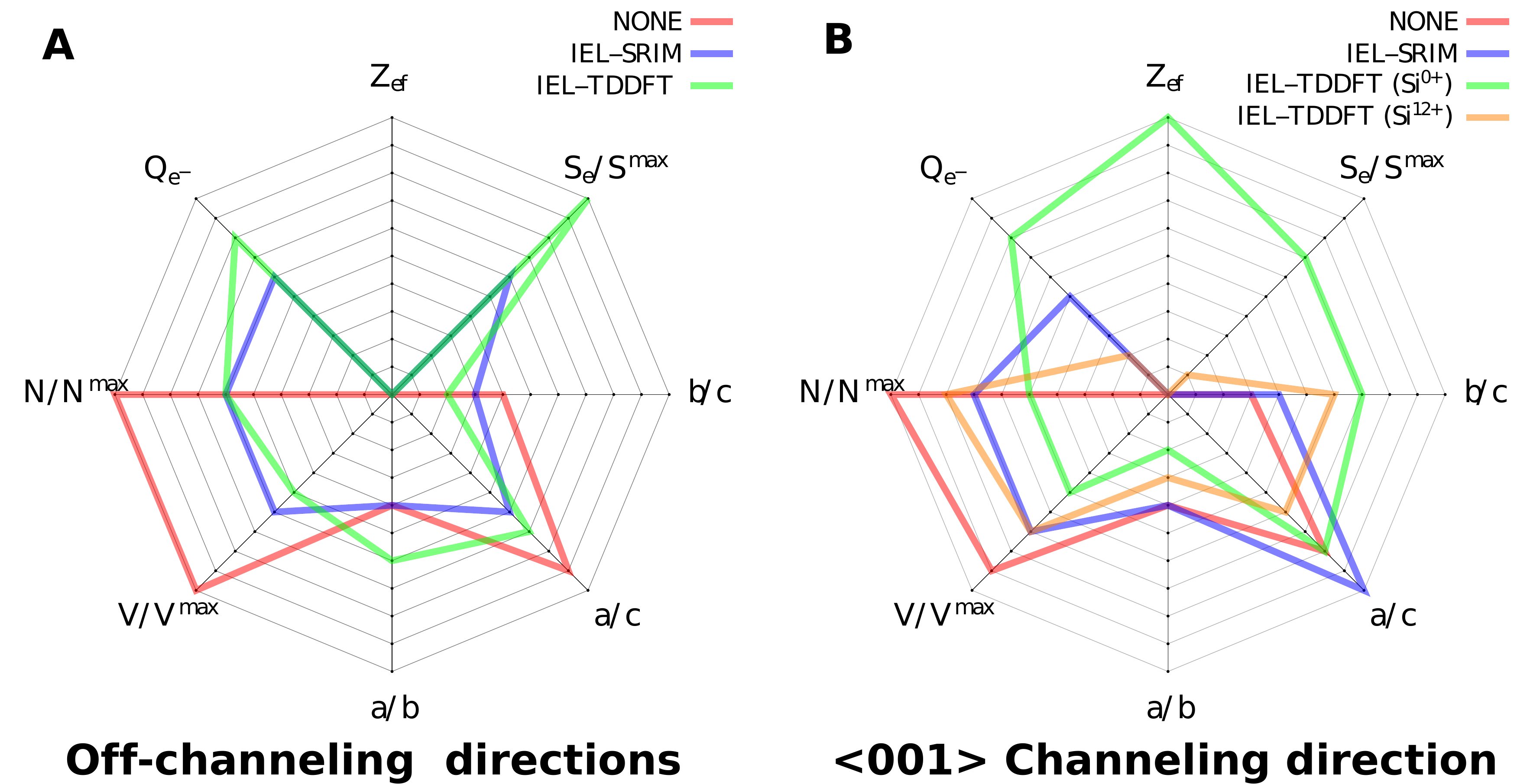}}
\caption{\label{fig:radar}
Comparison of cascade characteristics for (A) off-channeling direction, and (B) $\langle 100 \rangle$ channeling direction at 20 keV projectile kinetic energy.
Radar chart plots show normalized electronic stopping power $S_e/S^{\mathrm{max}}$ from RT-TDDFT, equilibrium/effective charge $Z_\mathrm{eff}$ (in e), total energy loss due to electronic stopping in MD simulations $Q_e^-$ (in keV), defect count $N/N^{\mathrm{max}}$ normalized with respect to result without electronic stopping, cascade volume $V/V^{\mathrm{max}}$ normalized with respect to result without electronic stopping, and cascade aspect ratios $a/b$, $a/c$, and $b/c$.
Red contour represents cascade characteristics when no electronic stopping is considered, blue contour is for electronic stopping parameterized by SRIM, green contour is for electronic stopping parameterized by RT-TDDFT for Si$^{+12}$, orange contour is for electronic stopping parameterized by RT-TDDFT for Si$^{+0}$.
}
\end{figure}

Subsequently, our detailed results for spatial, temporal, and thermal aspects of the damage cascade provide us with a means to estimate the residual damage surviving the quenching and annealing phase of the cascade development that will influence the micro-structure evolution.
The radar charts in Fig.\ \ref{fig:radar} graphically represent the differences in defect production and cascade morphology based on the representation of electronic stopping.
In this figure, the radial directions indicate the normalized stopping power at 20 keV $S_e/S^{max}$ with respect to the stopping power calculated using RT-TDDFT, the equilibrium charge at 20 keV ($Q$), the total energy loss due to electronic stopping during the simulation ($Q_{e}^-$), the normalized defect count with respect to the defect count when no electronic stopping is considered ($N/N^{max}$), the normalized cascade volume with respected to the volume of the cascade when no electronic stopping is considered, and the cascade aspect ratios ($a/b, a/c, b/c$).

In the case of the off-channeling direction, we note little differences in terms of the cascade structure ($a/b, a/c, b/c$ and $V/V^{max}$) when electronic stopping is represented either using SRIM or RT-TDDFT.
In contrast, as seen in Fig.\ \ref{fig:radar}B in the case of the channeling direction, the representation of the electronic stopping has a consequent impact on the defect production and cascade morphology.
Comparing the IEL-SRIM approximation with the IEL-TDDFT we note that not only the defect count and defect morphology are substantially different, but also that the charge state of the incident particle has a direct impact on the cascade characteristics.
This is also illustrated when comparing total energy loss and effective stopping power and is not surprising, since the charge state directly contributes to both of these quantities.

There are several practical implications of our work:
In particular, for ion-beam treatment of materials we envision targeted manipulation of properties of projectiles, such as charge or kinetic energy, based on our simulations.
Selection of the initial projectile kinetic energy (high vs.\ low) and impact angle (different channels and off-channeling) determines how the projectile subsequently interacts with the electrons of the target material and should allow for tuning of the final projectile charge state (see Fig.\ \ref{fig:KE_CHG_Si_Si}).
We show that since equilibration of highly ionized, channeling Si$^{+12}$ projectiles occurs through attracting valence electrons, their equilibrium charge state depends on the electron density distribution and binding energy of valence electrons, both of which depend on the specific target material.
Contrary, we show that the equilibrium charge state of neutral, channeling Si projectiles is exclusively determined by the electron loss of the projectile and, hence, should be independent of the target material.
Literature data to confirm these predictions is sparse and inconclusive:
A study \cite{Martin_1969} on a channeling oxygen ion with an initial charge that is smaller than the equilibrium charge found for off-channeling agrees with our prediction.
However, a study on iodine-irradiated gold targets reports that an initially weakly charged, channeling iodine ion loses fewer electrons than an off-channeling one \cite{Lutz_1970}.
Understanding the origin of these observations, e.g., by invoking the electronic structure of these materials, is an interesting area for future work.

Furthermore, accurate electronic-stopping measurements would be helpful to directly confirm our predictions for the equilibrium projectile charge:
Initially neutral and initially ionized slow projectiles equilibrate to different charge states when they move on a channel and their equilibration lengths also differ, depending on the initial charge state.
Hence, electronic stopping significantly differs for otherwise identical experimental conditions.
In particular, initially neutral channeling projectiles experience smaller electronic stopping, both due to the lower charge state and the different equilibration length.
Thus, they travel deeper into the target material than initially highly charged projectiles.
We also observe that initially ionized projectiles take longer to reach charge equilibrium when on a channel instead of off-channeling, due to the weaker interaction with semi-core electrons.
These effects allow experimental control over the ion range by selecting the initial projectile charge and trajectory.
At the same time, our MD simulations show that the projectiles that experience less electronic stopping create more defects and more extended cascades along their trajectory. 

Better experimental understanding of the pre-equilibrium stage would be particularly interesting.
Careful direct measurements of the charge state, e.g., for ions channeling through thin films or 2D materials, should allow for direct observation of the differences predicted from our simulations, including measurements of the thickness dependence of the projectile charge prior to equilibration.
As an example, Ref.\ \onlinecite{Wilhelm_2016, Gruber_2016} report a dependence of electronic stopping (electronic energy loss) on the initial and exit charge state of the projectile for slow (KE $\approx$ 6.8 keV) and highly charged ($Q>10$) Xe ions traveling through thin ($\approx$ nm) carbon membranes.
This experimental setup minimizes the opportunity for charge state to reach equilibrium and therefore works as a great direct probe to understand the effect of initial charge state on electronic stopping, in conjunction with precise simulations.
Both papers report that a larger initial charge gives rise to larger electronic stopping, essentially confirming our finding.
In addition, they report that a larger \emph{change} in charge state during the impact gives rise to larger electronic stopping and attribute this to the energy loss caused by the charge transfer.
However, based on our results, a larger change in charge state can result from a smaller impact parameter, which indicates larger local charge density for the projectile to interact with.
We conjecture that this larger local charge density also contribute to the larger observed electronic stopping.

\section{Conclusions}

We addressed the multi-scale nature of electron-ion dynamics in heavy-ion irradiated silicon by combining real-time time-dependent density functional theory and molecular dynamics based on the two-temperature model.
Our first-principles simulations reveal the detailed charge state dynamics of projectile ions and we explain the consequences on electronic stopping.
We show that electronic stopping of highly ionized Si projectiles on channeling trajectories is higher than for off-channeling ones across a wide kinetic energy range.
While this finding is opposite to what is expected for weakly ionized Si projectiles, we explain it by invoking the charge state of the projectile and find consistency with some of the previous experiments.
Furthermore, integration with full-cascade molecular dynamics simulations demonstrates the importance of understanding the detailed electron-ion dynamics during the impact. 
We show that different electronic stopping gives rise to qualitatively different cascade structures, which is critical for cascade simulations, e.g.\ for understanding ion-beam techniques and radiation damage.

\begin{acknowledgments}
The authors would like to thank K. Hattar from Sandia National Laboratories for insightful discussions during the execution of this work.
C.-W.L.\ acknowledges support from the Government Scholarship to Study Abroad from the Taiwan Ministry of Education.
C.-W.L.\ and A.S.\ gratefully acknowledge financial support from Sandia National Laboratories through the Sandia-UIUC collaboration (SNL grant No.\ 1736375) and from the Office of Naval Research (grant No.\ N00014-18-1-2605).
An award of computer time was provided by the Innovative and Novel Computational Impact on Theory and Experiment (INCITE) program.
This research used resources of the Argonne Leadership Computing Facility, which is a DOE Office of Science User Facility supported under Contract DE-AC02-06CH11357.
Support from the IAEA F11020 CRP "Ion Beam Induced Spatio-temporal Structural Evolution of Materials: Accelerators for a New Technology Era" is gratefully acknowledged.
This work was performed, in part, at the Center for Integrated Nanotechnologies, an Office of Science User Facility operated for the U.S. Department of Energy (DOE) Office of Science.
Sandia National Laboratories is a multi-mission laboratory managed and operated by National Technology and Engineering Solutions of Sandia, LLC., a wholly owned subsidiary of Honeywell International, Inc., for the U.S.\ DOE's National Nuclear Security Administration under contract DE-NA0003525.
The views expressed in the article do not necessarily represent the views of the U.S.\ DOE or the United States Government.
All essential data are presented in the paper and the Supplementary Materials at [URL will be
 inserted by publisher]. TDDFT results are available at the Materials Data Facility \cite{MDF,data}
\end{acknowledgments}

\bibliography{Si.bib}
\end{document}